\documentclass[11pt,twoside,letterpaper]{article} 
\usepackage{times,fancyhdr}
\usepackage[dvips]{graphicx}
\usepackage{amssymb}
\usepackage{latexsym}
\usepackage{amsmath}
\usepackage{float}


\makeatletter
\def\cleardoublepage{\clearpage\if@twoside \ifodd\c@page\else%
    \hbox{}%
    \thispagestyle{empty}%
    \newpage%
    \if@twocolumn\hbox{}\newpage\fi\fi\fi}
\makeatother

\renewcommand{\thesection}{\arabic{section}.}
\renewcommand{\thesubsection}{\thesection\arabic{subsection}.}

\def\figurename{Figure}
\makeatletter
\renewcommand{\fnum@figure}[1]{\figurename~\thefigure.}
\makeatother

\def\tablename{Table}
\makeatletter
\renewcommand{\fnum@table}[1]{\tablename~\thetable.}
\makeatother

\begin{document}
\title{
{\begin{flushleft}
\vskip 0.45in
{\normalsize\bfseries\textit{Chapter~1}}
\end{flushleft}
\vskip 0.45in
\bfseries\scshape Quantum Hall fluids in the presence of topological defects:
noncommutativity, generalized magnetic translations and non-Abelian
statistics}}
\author{\bfseries\itshape Patrizia Iacomino (1), Vincenzo Marotta (2)\thanks{E-mail address: v.marott@tiscali.it}, Adele Naddeo (3)\thanks{E-mail address: naddeo@sa.infn.it}\\
(1) Dipartimento di
Matematica e Applicazioni ''R. Caccioppoli'',\\
Universit\'{a} di Napoli ``Federico II'', \\
Compl.\ universitario M.
Sant'Angelo, Via Cinthia, 80126 Napoli, Italy\\
(2) Dipartimento di Scienze Fisiche, Universit\'{a} di Napoli ``Federico II''\\
and INFN, Sezione di Napoli, \\
Compl.\ universitario M. Sant'Angelo,
Via Cinthia, 80126 Napoli, Italy\\
(3) Dipartimento di Fisica \textit{''}E. R. Caianiello'',
Universit\'{a} degli Studi di Salerno,\\
and CNISM, Unit\'{a} di Ricerca di Salerno,\\
Via Ponte Don Melillo, 84084 Fisciano (SA), Italy}
\date{}
\maketitle
\thispagestyle{empty}
\setcounter{page}{1}
\thispagestyle{fancy}
\fancyhead{}
\fancyhead[L]{In: Book Title \\
Editor: Editor Name, pp. {\thepage-\pageref{lastpage-01}}} 
\fancyhead[R]{ISBN 0000000000  \\
\copyright~2007 Nova Science Publishers, Inc.}
\fancyfoot{}
\renewcommand{\headrulewidth}{0pt}

\vspace{2in}

\vspace{.20in}


\begin{abstract}
We review our recent results on the physics of quantum Hall fluids at Jain
and non conventional fillings within a general field theoretic framework. We
focus on a peculiar conformal field theory (CFT), the one obtained by means
of the $m$-reduction technique, and stress its power in describing strongly
correlated low dimensional condensed matter systems in the presence of
localized impurities or topological defects. By exploiting the notion of
Morita equivalence for field theories on noncommutative two-tori and
choosing rational values of the noncommutativity parameter, we find a
general one-to-one correspondence between the $m$-reduced conformal field
theory describing the quantum Hall fluid and an Abelian noncommutative field
theory. In this way we give a meaning to the concept of ''noncommutative
conformal field theory'', as the Morita equivalent version of a CFT defined
on an ordinary space. In this context the image of Morita duality in the
ordinary space is given by the $m$-reduction technique and the corresponding
noncommutative torus Lie algebra is naturally realized in terms of
generalized magnetic translations.

As an example of application of the formalism, we study a quantum Hall
bilayer at nonconventional fillings in the presence of a localized
topological defect and briefly recall its boundary state structure
corresponding to two different boundary conditions, the periodic as well as
the twisted boundary conditions respectively, which give rise to different
topological sectors on a torus. By analyzing the boundary interaction terms
present in the action we recognize a boundary magnetic term and a boundary
potential. Then we introduce generalized magnetic translation operators as
tensor products, which act on the quantum Hall fluid and defect space
respectively, and compute their action on the boundary partition functions:
in this way their role as boundary condition changing operators is fully
evidenced. From such results we infer the general structure of generalized
magnetic translations in our model and clarify the deep relation between
noncommutativity and non-Abelian statistics of quasi-hole excitations, which
is crucial for physical implementations of topological quantum computing. In
particular, noncommutativity is strictly related to the presence of a
topological defect on the edge of the bilayer system, which supports
protected Majorana fermion zero modes. That happens in close analogy with
point defects in topological insulators and superconductors, where the
existence of Majorana bound states is related to a $Z_{2}$ topological
invariant. Finally, some prospects on the implementation of a topologically
protected qubit with quantum Hall bilayers are presented.
\end{abstract}


\pagestyle{fancy}
\fancyhead{}
\fancyhead[EC]{Patrizia Iacomino, Vincenzo Marotta, Adele Naddeo}
\fancyhead[EL,OR]{\thepage}
\fancyhead[OC]{Quantum Hall fluids}
\fancyfoot{}
\renewcommand\headrulewidth{0.5pt}

\raggedbottom                         
\section{Introduction}

The experimental discovery of fractional quantum Hall (FQH) effect \cite
{hallexp1} in 1982 has opened the door to a new fascinating state of matter.
Indeed the unusual properties of the incompressible quantum fluid that
arises in a two dimensional electron gas subjected to a strong magnetic
field and at very low temperatures are the signature of an emergent
topological state of matter, whose quasiparticle excitations show up
fractional charge and statistics \cite{prangegirvin}\cite{sarma}. As such,
it has been proposed that FQH states display a new kind of order, termed
topological order \cite{wen}\cite{cristofano1}. In particular they lack long
range correlation and local order parameters but display a weak form of
order which is sensitive to the topology of the underlying two dimensional
manifold. Further appealing features are a non-Abelian Berry's phase under
modular transformations and a ground state degeneracy depending on the
topology of the underlying space, which is robust against any local
perturbations.

Laughlin states \cite{laughlin1} with $\nu =\frac{1}{k}$ and $k$ odd are
today well understood, both theoretically and experimentally. In order to
take into account more general observed filling factors $\nu $ different
from $\frac{1}{k}$, a hierarchical scheme was introduced \cite{hier1}, in
which quasiparticles of a $\nu =\frac{1}{k}$ state can themselves condense
into a new quantized state. In this way it has been possible to construct
Hall states for any odd denominator filling fraction $\nu $, whose
quasiparticles have fractional charge and Abelian fractional statistics.
Some years later Jain introduced the composite fermions picture \cite{jain}
in order to explain more general filling fractions of the form $\nu =\frac{p%
}{2p+1}$. The observation of a quantum Hall state with an even denominator
filling fraction \cite{hallexp2}, $\nu =\frac{5}{2}$, paved the way to the
study of states which do not follow the hierarchical picture and then are an
exception to the odd denominator rule. The most promising theoretical
candidate for such a state is believed to be the Moore-Read (MR) Pfaffian
model \cite{MR}, whose anyonic quasiparticles exhibit non Abelian braiding
statistics. This last prediction is appealing in view of condensed matter
implementations of topological quantum computation \cite{tqc1}. Indeed
quantum information is stored in topologically degenerate states with
multiple quasiparticles while unitary gate operations are carried out by
braiding such quasiparticles and then by reading the corresponding states.
The topological nature of these quasiparticles states makes them robust
against any local perturbation. A strong support to the Moore-Read hypothesis
for the $\nu =\frac{5}{2}$ state comes from recent experimental data about
the charge $\frac{e}{4}$ of localized excitations \cite{hallexp3}, which
coincide with previous shot noise \cite{hallexp4} and quasiparticle
interference oscillation \cite{hallexp5} results. Further experimental
evidence has been gained through the observation of the predicted neutral
mode \cite{hallexp6}, also consistent with the MR picture.

At the same time, increasing technological progress in molecular beam
epitaxy techniques has led to the ability to produce pairs of closely spaced
two-dimensional electron gases. Since then such bilayer quantum Hall systems
have been widely investigated theoretically as well as experimentally \cite
{sarma}\cite{teoria, eisenstein}. Strong correlations between the electrons
in different layers lead to new physical phenomena involving spontaneous
interlayer phase coherence with an associated Goldstone mode. In particular
a spontaneously broken $U(1)$ symmetry \cite{Zee} has been discovered and
identified and many interesting properties of such systems have been
studied: the Kosterlitz-Thouless transition, the zero resistivity in the
counter-current flow, a DC/AC Josephson-like effect in interlayer tunneling
as well as the presence of a gapless superfluid mode \cite{Zee1}\cite
{girvin1}. Indeed, when tunneling between the layers is weak, the quantum
Hall bilayer state can be viewed as arising from the condensation of an
excitonic superfluid in which an electron in one layer is paired with a hole
in the other layer. The uncertainty principle makes it impossible to tell
which layer either component of this composite boson is in. Equivalently the
system may be regarded as a ferromagnet in which each electron exists in a
coherent superposition of the ''pseudospin'' eigenstates, which encode the
layer degrees of freedom \cite{YangMoon}\cite{girvin1}. The phase variable
of such a superposition fixes the orientation of the pseudospin magnetic
moment and its spatial variations govern the low energy excitations in the
system. So quantum Hall bilayers are an interesting realization of the
pairing picture at non-standard fillings. Indeed they show up several even
denominator states.

The topological nature of quantum Hall states makes possible the
classification of the different electronic phases of the quantum liquid
according to topological invariants, as pioneered by Thouless \cite{thouless}%
. He identified the integer topological invariant characterizing the two
dimensional ($2D$) integer quantum Hall state, which gives the Hall
conductance and characterizes the Bloch Hamiltonian defined in the magnetic
Brillouin zone. As a consequence of this topological classification a
bulk-boundary correspondence arises, which relates the topological class of
the bulk system to the number of gapless chiral edge states on the sample
boundary \cite{wen}. Topologically protected zero modes and gapless states
can also occur when topological defects are present on the edge of the
sample; recently a generalization of the bulk-boundary correspondence has
been introduced as well \cite{tk1}, which relates the topological class of
the Hamiltonian characterizing the defect to the structure of the protected
modes associated with the defect. In this way the crucial role of localized
topological defects clearly emerges and that appears to be deeply related to
noncommutativity, as we showed in our recent work \cite{AV1}\cite{AV2}\cite
{AV3}.

On the other hand, the relevance of $2D$ CFT for the fractional quantum Hall
effect (FQHE) was first pointed out in Ref. \cite{fubini1}, where the
analogy between the Laughlin-Jastrow (LJ) wave function and the dual
amplitudes was exploited. Then the $2D$ CFT formalism was extensively
employed in the study of Hall fluids at the plateaux, assuming that it can
describe successfully the universal topological features of the fractional
quantum Hall effect (FQHE) \cite{napoli1}\cite{cristofano1}. The key point
for the introduction of $2D$ CFT is the consideration that it is possible to
build up a Coulomb Gas in terms of vertex operators of a $c=1$ CFT and then
to compute in a natural way topological quantities such as the Hall
conductance. That relies strongly on the one-component plasma interpretation
of the modulus square of the Hall ground state wave function proposed by
Laughlin \cite{laughlin1}. Furthermore the vertex operator formalism
describes very well states with particles carrying an electric as well as a
magnetic charge, i. e. dyons. We also point out that the $2D$ CFT approach
to FQHE holds not only on the plane but also on a torus \cite{torus1} and in
general on a Riemann surface with arbitrary genus; in this way the
topological properties of the system can be made very transparent. Indeed
the topological nature of the order present in the FQHE at fillings $\nu =%
\frac{1}{k}$ shows up as a $k$-fold degeneracy of the ground state wave
function. Such a result, as well as the value of the Hall conductance, is
deeply related \cite{cristofano1} to the algebraic properties of a finite
subgroup of the magnetic translation group for a $2D$ electrons system. More
recently, a particular CFT, the one obtained via $m$-reduction technique
\cite{VM}, has been introduced by our group and applied to the description
of a quantum Hall fluid (QHF) at Jain \cite{cgm1}\cite{cgm3} as well as
paired states fillings \cite{cgm2}\cite{cgm4} and in the presence of
topological defects \cite{noi1}\cite{noi2}\cite{noi5}. The $m$-reduction
technique is based on the simple observation that, for any CFT (mother), a
class of sub-theories exists, which is parameterized by an integer $m$ with
the same symmetry but different representations. The resulting theory
(daughter), called Twisted Model (TM), has the same algebraic structure but
a different central charge $c_{m}=mc$. Its application to the physics of the
QHF arises by the incompressibility of the Hall fluid droplet at the
plateaux, which implies its invariance under the $W_{1+\infty }$ algebra at
different fillings \cite{ctz5}, and by the peculiarity of the $m$-reduction
procedure to provide a daughter CFT with the same $W_{1+\infty }$ invariance
property of the mother theory \cite{cgm1}\cite{cgm3}. Thus the $m$-reduction
furnishes automatically a mapping between different incompressible plateaux
of the QHF while the characteristics of the daughter theory is the presence
of twisted boundary conditions on the fundamental fields.

But how noncommutativity does arise in the physics of quantum Hall regime
and how does it fit to our CFT description? Really, it comes out in a very
natural way as an effective description of the dynamics when the simplest
framework is considered, namely the quantum mechanics of the motion of $N_{e}
$ charged particles in two dimensions subjected to a transverse magnetic
field (Landau problem)\cite{landau}. The strong field limit $B\rightarrow
\infty $ at fixed mass $m$ projects the system onto the lowest Landau level
and, for each particle $I=1,...,N_{e}$, the corresponding coordinates ($%
\frac{eB}{c}x_{I},y_{I}$) are a pair of canonical variables which satisfy
the commutation relations $\left[ x_{I},y_{I}\right] =i\delta _{I,J}\theta $%
, $\theta =\frac{\hbar c}{eB}\equiv l_{M}^{2}$ being the noncommutativity
parameter. In this picture the electron is not a point-like particle and can
be localized at best at the scale of the magnetic length $l_{M}$. The same
thing happens to the endpoints of an open string attached to a $D$-brane
embedded in a constant magnetic field \cite{openstring1}, which is the
string theory analogue of the Landau problem: the $D$-brane worldvolume
becomes a noncommutative manifold. Indeed the worldsheet field theory for
open strings attached to $D$-branes is defined by a $\sigma $-model on the
string worldsheet $\Sigma $ with action
\begin{equation}
S_{\Sigma }=\frac{1}{4\pi l_{s}^{2}}\int_{\Sigma }d^{2}\xi \left(
g_{ij}\partial ^{a}y^{i}\partial _{a}y^{j}-2\pi \mathrm{i}%
l_{s}^{2}B_{ij}\epsilon ^{ab}\partial _{a}y^{i}\partial _{b}y^{j}\right) .
\label{mb1}
\end{equation}
Here $y^{i}$ are the open string endpoint coordinates, $\xi ^{a}$, $a=1,2$
are local coordinates on the surface $\Sigma $, $l_{s}$ is the intrinsic
string length, $g_{ij}$ is the spacetime metric and $B_{ij}$ is the
Neveu-Schwarz two-form which is assumed non-degenerate and can be viewed as
a magnetic field on the $D$-brane. Indeed, when $B_{ij}$ are constant the
second term in Eq. (\ref{mb1}) can be integrated by parts and gives rise to
the boundary action:
\begin{equation}
S_{\partial \Sigma }=-\frac{\mathrm{i}}{2}\int_{\partial \Sigma
}dtB_{ij}y^{i}\left( t\right) \overset{.}{y}^{j}\left( t\right) ,
\label{mb2}
\end{equation}
where $t$ is the coordinate of the boundary $\partial \Sigma $ of the string
world sheet lying on the $D$-brane worldvolume and $\overset{.}{y}%
^{i}=\partial y^{i}/\partial t$. Such a boundary action formally coincides
with the one of the Landau problem in a strong field. Now, by taking the
Seiberg-Witten limit \cite{witten1}, i. e. by taking $g_{ij}\sim
l_{s}^{4}\sim \varepsilon \rightarrow 0$ while keeping fixed the field $%
B_{ij}$, the effective worldsheet field theory reduces to the boundary
action (\ref{mb2}) and the canonical quantization procedure gives the
commutation relations $\left[ y^{i},y^{j}\right] =i\theta ^{ij}$, $\theta =%
\frac{1}{B}$ on $\partial \Sigma $. Summarizing, the quantization of the
open string endpoint coordinates $y^{i}\left( t\right) $ induces a
noncommutative geometry on the $D$-brane worldvolume and the effective
low-energy field theory is a noncommutative field theory (NCFT) \cite{ncft1}
for the massless open string modes.

In such a picture also a tachyon condensation phenomenon can be considered,
which introduces the following boundary interaction for the open string:
\begin{equation}
S_{T}=-\mathrm{i}\int_{\partial \Sigma }dtT\left( y^{i}\left( t,0\right)
\right) ,  \label{mb3}
\end{equation}
where $T\left( y^{i}\left( t,0\right) \right) $ is a general tachyon
profile. In general, $D$-branes in string theory correspond to conformal
boundary states, i. e. to conformally invariant boundary conditions of the
associated CFT. They are charged and massive objects that interact with
other objects in the bulk, for instance through exchange of higher closed
string modes. In turn boundary excitations are described by fields that can
be inserted only at points along the boundary, i. e. the boundary fields.
There exists an infinite number of open string modes, which correspond to
boundary fields in the associated boundary CFT. In other words tachyonic
condensation is a boundary phenomenon and as such it is related to
Kondo-like effects in condensed matter systems \cite{kondo1}\cite{affleck1}.
In this case, the presence of the background $B$-field makes the open string
states to disappear and that explains the independence on the background
\cite{seiberg1}.

The above picture coincides with the one proposed in Ref. \cite{maldacena}
in the context of boundary conformal field theories (BCFT). The authors
consider a system of two massless free scalar fields which have a boundary
interaction with a periodic potential and furthermore are coupled to each
other through a boundary magnetic term, whose expressions coincide with Eqs.
(\ref{mb2}) and (\ref{mb3}). By using a string analogy, the boundary
magnetic interaction allows for exchange of momentum of the open string
moving in an external magnetic field. Indeed it enhances one chirality with
respect to the other producing the effect of a rotation together with a
scale transformation on the fields and the string parameter plays the role
of dissipation. It is crucial to observe that conformal invariance of the
theory is preserved only at special values of the parameters entering the
action, the so called ``magic''\ points. That happens in close analogy with
the motion of an electron confined in a plane, subjected to an external
magnetic field $B$, normal to the plane, and in the presence of dissipation
\cite{cgm5}. Furthermore we have shown how our $m$-reduced theory at paired
states fillings describes a dissipative system precisely at the ``magic''\
points \cite{noi2}.

In this way noncommutativity comes into play and a deep relation emerges
between the quantum mechanical and the string and $D$-brane description of
the quantum Hall regime. Now, in order to show how it relates to our CFT
description of QHF, we start by considering quantum field theories defined on
a noncommutative two-torus and then resort to the concept of Morita duality
\cite{morita}\cite{scw1}\cite{scw2}. Such a kind of duality establishes a
relation, via a one-to-one correspondence, between representations of two
noncommutative algebras and, within the context of gauge theories on
noncommutative tori, it can be viewed as a low energy analogue of $T$%
-duality of the underlying string model \cite{string1}: as such, it results
a powerful tool in order to establish a correspondence between NCFT and well
known standard field theories. Indeed, for rational values of the
noncommutativity parameter, $\theta =\frac{1}{N}$, one of the theories
obtained by using the Morita equivalence is a commutative field theory of
matrix valued fields with twisted boundary conditions and magnetic flux $c$
\cite{wilson1} (which, in a string description, behaves as a $B$-field
modulus). Our recent work \cite{AV1}\cite{AV2}\cite{AV3} strongly relies on
such an idea. It aims at building up a general effective theory for QHF
which could add further evidence to the relationship between the string
theory picture and the condensed matter theory one as well as to the role of
noncommutativity in QHF physics. In particular, we show by means of the
Morita equivalence that a NCFT with $\theta =2p+\frac{1}{m}$ or $\theta =%
\frac{p}{2}+\frac{1}{m}$ respectively is mapped to a CFT on an ordinary
space. We identify such a CFT with the $m$-reduced CFT developed for a QHF
at Jain $\nu =\frac{m}{2pm+1}$ \cite{cgm1}\cite{cgm3}, as well as paired
states fillings $\nu =\frac{m}{pm+2}$ \cite{cgm2}\cite{cgm4}, whose neutral
fields satisfy twisted boundary conditions. Indeed the presence of a $Z_{m}$
twist is the fingerprint of a topological defect \cite{noi1}\cite{noi2}\cite
{noi5} localized somewhere on the edge of the system and accounts, in the
open string picture, for a mismatch of momentum exchange at its two
endpoints. In this way we give a meaning to the concept of ''noncommutative
conformal field theory'', as the Morita equivalent version of a CFT defined
on an ordinary space. The image of Morita duality in the ordinary space is
given by the $m$-reduction technique and the corresponding noncommutative
torus Lie algebra is naturally realized in terms of Generalized Magnetic
Translations (GMT). That introduces a new relationship between
noncommutative spaces and QHF and paves the way for further investigations
on the role of noncommutativity in the physics of general strongly
correlated many body systems \cite{ncmanybody}.

In this chapter we illustrate such developments and then, as an example
of application, we study in detail the physics of a system of two parallel
layers of $2D$ electrons gas in a strong perpendicular magnetic field and
interacting with an impurity placed somewhere on the boundary (quantum Hall
bilayer). We focus on the case of non standard filling factor, which is
relevant both from an experimental point of view and for possible
topological quantum computing implementations. Indeed it is described by a $m
$-reduced CFT with $m=2$ and we find that the $2$-reduced theory on the
two-torus obtained as a Morita dual starting from a NCFT keeps track of
noncommutativity in its structure. Furthermore GMT are a realization of the
noncommutative torus Lie algebra. We analyze in detail the presence of a
topological defect placed between the layers somewhere on the edges and
discuss the relation between different defects and different possible
boundary conditions by introducing the corresponding boundary partition
function. A boundary state can be defined in correspondence to each class of
defects \cite{noi1} and a boundary partition function can be computed which
corresponds to a boundary fixed point, e. g. to a different topological
sector of the theory on the torus. In this context GMT are identified with
operators which act on the boundary states and realize the transition
between fixed points of the boundary flow. In the language of Kondo effect
\cite{affleck1} they behave as boundary condition changing operators. We
introduce general GMT operators as tensor products which act on the QHF and
defect space respectively and discuss in detail their behaviour. Then, the
emergence of noncommutativity as a consequence of the presence of the
topological defect is emphasized and its connection with non-Abelian
statistics of the quasi-hole excitations fully elucidated. We find for such
excitations a $SO\left( 2n\right) $ structure, typical of Ising anyons \cite
{nayak}\cite{isinganyons}, while the topological defect supports protected
Majorana fermion zero modes in close analogy with point defects in
topological insulators and superconductors \cite{tk1}\cite{tk2}. Finally we
give some insights on how to build up a topologically protected qubit with a
quantum Hall bilayer when two localized defects are introduced on the edge
\cite{AV4}.

The outline of the chapter is the following.

In Section 2, we give a brief account of our theoretical approach, the $m$%
-reduction procedure \cite{VM}, and illustrate how it works in the
description of a QHF at Jain as well as paired states fillings. In this last
case we discuss explicitly the $m=2$ theory, which corresponds to a quantum
Hall bilayer in the presence of a localized topological defect.

Section 3 is devoted to show how noncommutativity comes into play in our CFT
description by focusing on the issue of Morita equivalence for field
theories on noncommutative two-tori. That allows us to introduce a new
relationship between noncommutative spaces and QHF, which is explicitly
discussed for Jain $\nu =\frac{m}{2pm+1}$ \cite{cgm1}\cite{cgm3}, as well as
paired states fillings $\nu =\frac{m}{pm+2}$ \cite{cgm2}\cite{cgm4}. In this
last case we make explicit reference to the bilayer theory $m=2$, which will
be the subject of our study in the following Sections.

In Section 4 we focus on the physics of a quantum Hall bilayer at paired
states fillings because it is the simplest non trivial example on which all
the relevant features of our theory can be exploited. We discuss in detail
the different possible boundary interactions of the system and show how it
is equivalent to a system of two massless scalar bosons with a magnetic
boundary interaction at particular points \cite{maldacena}. Then the
boundary content of our theory is rephrased in terms of boundary partition
functions, which are shown to be closed upon action of GMT. These results
allow us to infer the structure of the most general GMT operators.

In Section 5 we present and discuss in detail the structure of GMT for
quantum Hall bilayers as tensor products acting on the QHF and defect space
respectively. In particular we find an interesting relation between
noncommutativity and non-Abelian statistics of quasihole excitations, as a
consequence of the presence of a defect. Finally we briefly sketch a
possible implementation of a topologically protected qubit with two localized
defects introduced on the edge of the bilayer system.

In Section 6, some comments and outlooks of our work are given.

Finally, the operator content of our theory, the TM, on the torus for a
quantum Hall bilayer at paired states fillings is recalled in the Appendix.

\section{The $m$-reduction technique}

In this Section we briefly review the basics of the $m$-reduction procedure
on the plane (genus $g=0$) \cite{VM} and then we show how it works,
referring to the description of a QHF at Jain $\nu =\frac{m}{2pm+1}$ \cite
{cgm1}\cite{cgm3} as well as paired states fillings $\nu =\frac{m}{pm+2}$
\cite{cgm2}\cite{cgm4}.

In general, the $m$-reduction technique is based on the simple observation
that for any CFT (mother) exists a class of sub-theories parameterized by an
integer $m$ with the same symmetry but different representations. The
resulting theory (daughter) has the same algebraic structure but a different
central charge $c_{m}=mc$. In order to obtain the generators of the algebra
in the new theory we need to extract the modes which are divided by the
integer $m$. These can be used to reconstruct the primary fields of the
daughter CFT. This technique can be generalized and applied to any extended
chiral algebra which includes the Virasoro one. Following this line one can
generate a large class of CFTs with the same extended symmetry but with
different central extensions. It can be applied in particular to describe
the full class of Wess-Zumino-Witten (WZW) models with symmetry $\widehat{%
su(2)}_{m}$, obtaining the associated parafermions in a natural way or the
incompressible $W_{1+\infty }$ minimal models \cite{ctz5} with central
charge $c=m$. Indeed the $m$-reduction preserves the commutation relations
between the algebra generators but modify the central extension (i.e. the
level for the WZW models). In particular this implies that the number of the
primary fields gets modified.

The general characteristics of the daughter theory is the presence of
twisted boundary conditions (TBC) which are induced on the component fields
and are the signature of an interaction with a localized topological defect.
It is illuminating to give a geometric interpretation of that in terms of
the covering on a $m$-sheeted surface or complex curve with branch-cuts, see
for instance Figs. 1, 2 for the particular case $m=2$.

\begin{figure}[h]
\centering\includegraphics*[width=0.3\linewidth]{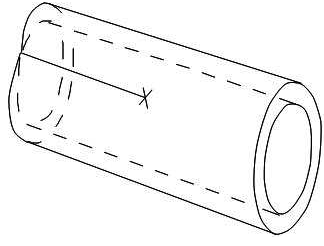}
\caption{The edge of the $2$-covered cylinder can be viewed as a separation
line of two different domains of the $2$-reduced CFT.}
\label{figura1}
\end{figure}

Indeed the fields which are defined on the left domain of the boundary have
TBC while the fields defined on the right one have periodic boundary
conditions (PBC). When we generalize the construction to a Riemann surface
of genus $g=1$, i. e. a torus, we find different sectors corresponding to
different boundary conditions on the cylinder, as shown in detail in Refs.
\cite{cgm3}\cite{cgm4}. Finally we recognize the daughter theory as an
orbifold of the usual CFT describing the QHF at a given plateau.

\begin{figure}[h]
\centering\includegraphics*[width=0.3\linewidth]{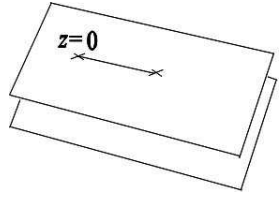}
\caption{The local branched plane.}
\end{figure}

The physical interpretation of such a construction within the context of a
QHF description is the following. The two sheets simulate a two-layer
quantum Hall system and the branch cut represents TBC which emerge from the
interaction with a localized topological defect on the edge \cite{noi1}\cite
{noi2}\cite{noi5}.

Let us now briefly summarize our $m$-reduction procedure on the plane \cite
{VM}. The starting point is described by a CFT with $c=1$, in terms of a
scalar chiral field compactified on a circle with general radius $R^{2}$ ($%
R^{2}=1$ for the Jain series \cite{cgm1} while $R^{2}=2$ for the non
standard one \cite{cgm2}, as we will recall in the following). Then the $%
u(1) $ current is given by $J(z)=i\partial _{z}Q(z)$, where $Q(z)$ is the
compactified Fubini field with the standard mode expansion:
\begin{equation}
Q(z)=q-i\,p\,lnz+\sum_{n\neq 0}\frac{a_{n}}{n}z^{-n},  \label{modes}
\end{equation}
where $a_{n}$, $q$ and $p$ satisfy the commutation relations $\left[
a_{n},a_{n^{\prime }}\right] =n\delta _{n,n^{\prime }}$ and $\left[ q,p%
\right] =i$. The primary fields are expressed in terms of the vertex
operators $U^{\alpha _{s}}(z)=:e^{i\alpha _{s}Q(z)}:$ with $\alpha _{s}=%
\frac{s}{R}$\ ($s=1,...,R^{2}$) and conformal dimension $h=\frac{s^{2}}{%
2R^{2}}$.

Starting with the set of fields in the above CFT and using the $m$-reduction
procedure, which consists in considering the subalgebra generated only by
the modes in Eq. (\ref{modes}), which are multiple of an integer $m$, we get
the image of the twisted sector of a $c=m$ orbifold CFT (i. e. the TM),
which describes the Lowest Landau Level (LLL) dynamics of the new filling in
the QHF context. In this way the fundamental fields are mapped into $m$
twisted fields which are related by a discrete Abelian group. Indeed the
fields in the mother CFT can be factorized into irreducible orbits of the
discrete $Z_{m}$ group, which is a symmetry of the TM, and can be organized
into components, which have well defined transformation properties under
this group. To compare the orbifold so built with the $c=m$ CFT, we use the
mapping $z\rightarrow z^{1/m}$ and the isomorphism, defined in Ref. \cite{VM}%
, between fields on the $z$ plane and fields on the $z^{m}$ covering plane
given by the following identifications: $a_{nm+l}\longrightarrow \sqrt{m}%
a_{n+l/m}$, $q\longrightarrow \frac{1}{\sqrt{m}}q$.

We perform a \textquotedblleft double\textquotedblright\ $m$-reduction which
consists in applying this technique into two steps.

\textbf{1)} The $m$-reduction is applied to the Fubini field $Q(z).$ That
induces twisted boundary conditions on the currents. It is useful to define
the invariant scalar field:
\begin{equation}
X(z)=\frac{1}{m}\sum_{j=1}^{m}Q(\varepsilon ^{j}z),  \label{X}
\end{equation}
where $\varepsilon ^{j}=e^{i\frac{2\pi j}{m}}$, corresponding to a
compactified boson on a circle with radius now equal to $R_{X}^{2}=R^{2}/m$.
This field describes the $U(1)$ electrically charged component of the new
filling in a QHF description.

On the other hand the non-invariant fields defined by
\begin{equation}
\phi ^{j}(z)=Q(\varepsilon ^{j}z)-X(z),~~~~~~~~~~~~~~~~\sum_{j=1}^{m}\phi
^{j}(z)=0  \label{phi}
\end{equation}
naturally satisfy twisted boundary conditions, so that the $J(z)$ current of
the mother theory decomposes into a charged current given by $J(z)=i\partial
_{z}X(z)$ and $m-1$ neutral ones $\partial _{z}\phi ^{j}(z)$.

\textbf{2)} The $m$-reduction applied to the vertex operators $U^{\alpha
_{s}}(z)$ of the mother theory also induces twisted boundary conditions on
the vertex operators of the daughter CFT. The discrete group used in this
case is just the $m$-ality group which selects the neutral modes with a
complementary cut singularity, which is necessary to reinforce the locality
constraint.

The vertex operator in the mother theory can be factorized into a vertex
that depends only on the $X(z)$ field:
\begin{equation}
\mathcal{U}^{\alpha _{s}}(z)=z^{\frac{\alpha _{s}^{2}(m-1)}{m}}:e^{i\alpha
_{s}{\ }X(z)}:
\end{equation}
and in vertex operators depending on the $\phi ^{j}(z)$ fields. It is useful
to introduce the neutral component:
\begin{equation}
\psi _{1}(z)=\frac{z^{\frac{1-m}{m}}}{m}\sum_{j=1}^{m}\varepsilon
^{j}:e^{i\phi ^{j}(z)}:
\end{equation}
which is invariant under the twist group given in \textbf{1)} and has $m$%
-ality charge $l=1$. Then, the new primary fields are the composite vertex
operators $V^{\alpha _{s}}(z)=\mathcal{U}^{\alpha _{s}}(z)\psi _{l}(z)$,
where $\psi _{l}$ are the neutral operators with $m$-ality charge $l$.

\bigskip From these primary fields we can obtain the new Virasoro algebra
with central charge $c=m$ which is generated by the energy-momentum tensor $%
T(z)$. It is the sum of two independent operators, one depending on the
charged sector:
\begin{equation}
T_{X}(z)=-\frac{1}{2}:\left( \partial _{z}X\left( z\right) \right) ^{2}:
\label{VIR1}
\end{equation}
with $c=1$ and the other given in terms of the $Z_{m}$ twisted bosons $\phi
^{j}(z)$:
\begin{equation}
T_{\phi }(z)=-\frac{1}{2}\sum_{j,j^{\prime }=1}^{m}:\partial _{z}\phi
^{j}(z)\partial _{z}\phi ^{j^{\prime }}(z):+\ \frac{m^{2}-1}{24mz^{2}}
\label{VIR2}
\end{equation}
with $c=m-1$.

Let us notice here that, although the daughter CFT has the same central
charge value, it differs in the symmetry properties and in the spectrum,
depending on the mother theory we are considering, i.e. for Jain \cite{cgm1}
or non standard series \cite{cgm2} in the case of a QHF as we will show in
the following.

\subsection{Jain fillings}

In this Subsection we focus on the description of a QHF at Jain fillings in
terms of vertex operators and review the main results of $m$-reduction
procedure in order to classify its excitations. The starting point is a CFT
with $c=1$, in terms of a scalar chiral field compactified on a circle with
radius $R^{2}=1$. Then the $U(1)$ current is given by $J(z)=i\partial
_{z}Q(z)$, where $Q(z)$ is the compactified Fubini field given in Eq. (\ref
{modes}). The primary fields are expressed in terms of the vertex operators $%
U^{\alpha _{s}}(z)=:e^{i\alpha _{s}Q(z)}:$ with $s=1$ and conformal
dimension $h=\frac{1}{2}$. The dynamical symmetry is given by the $%
W_{1+\infty }$ algebra \cite{BS} with $c=1$, whose generators are simply
given by a power of the current $J(z)$. By using the $m$-reduction
procedure, we get the image of the twisted sector of a $c=m$ orbifold CFT
which has $\widehat{U}(1){\times }\widehat{SU}(m)_{1}$ as extended symmetry
and describes the QHF at the new general filling $\nu =\frac{m}{2pm+1}$. In
order to do so, we factorize the fields into two parts, the first is the $%
c_{X}=1$ charged sector with radius $R_{X}^{2}=\frac{2pm+1}{m}$, the second
describes neutral excitations with total conformal central charge $c_{\phi
}=m-1$ for any $p\in N$ \cite{cgm1}.

In order to obtain a pure holomorphic wave function we have to consider the
correlator of the TM primary fields, which are the composite vertex
operators $V^{\alpha _{s}}(z)=\mathcal{U}^{\alpha _{s}}(z)\psi _{l}(z)$
\footnote{$\psi _{l}$ are the neutral operators associated with
representations of $m$-ality $l$ of \ $\widehat{SU}(m)_{1}$\cite{cgm1}.}
with conformal dimension:
\begin{equation}
h_{l}=\frac{l^{2}}{2m\left( 2pm+1\right) }+\frac{a}{2}\left( \frac{m-a}{m}%
\right) ,\text{ \ \ \ \ \ \ \ \ }l=1,2,...,m\left( 2pm+1\right) ;
\label{cdim1}
\end{equation}
they describe excitations with electric charge $q_{e}=\frac{l}{2pm+1}$ and
magnetic charge $q_{m}=l$ in units of $\frac{hc}{e}$. There exist also
integer charge quasi-particles (termed $a$-electrons), with half integer (or
integer) conformal dimension given by:
\begin{equation}
h_{l}=a^{2}p+\frac{a}{2},\text{ \ \ \ \ \ \ \ \ }l=\left( 2pm+1\right) a;%
\text{ \ \ \ \ \ \ \ }a=1,2,...,m.  \label{cdim2}
\end{equation}
In particular the electrons are obtained in correspondence of $q_{e}=1$ and $%
q_{m}=2pm+1$, while the other $2pm$ primary fields correspond to anyons.

The spectrum just obtained follows from the construction of the Virasoro
algebra with central charge $c=m$ (see Eqs. (\ref{VIR1})-(\ref{VIR2})). We
should point out that $m$-ality in the neutral sector is coupled to the
charged one exactly, as it was derived in Refs. \cite{frohlich}\cite{ctz5}
according to the physical request of locality of the electrons with respect
to the edge excitations. Indeed our projection, when applied to a local
field (namely the electron field in the case of filling factor $\nu =1$),
automatically couples the discrete $Z_{m}$ charge of $U(1)$ with the neutral
sector in order to give rise to a well defined, i. e. single valued,
composite field. Let us also notice that the $m$-electron vertex operator
does not contain any neutral field, so its wave function is realized only by
means of the $c_{X}=1$ charged sector: we deal with a pseudoparticle with
electric charge $m$ and magnetic charge $2pm+1$. The above construction has
been generalized to the torus topology as well \cite{cgm3}, confirming the
picture just outlined for the spectrum of excitations of a QHF at Jain
fillings.

\subsection{Paired states fillings}

Let us now review how the $m$-reduced theory describes successfully a QHF at
paired states fillings $\nu =\frac{m}{pm+2}$ \cite{cgm2,cgm4}. We focus
mainly on the special case $m=2$ and on the physics of a quantum Hall
bilayer, which will be of our interest in the following sections as a case
study to illustrate the main theoretical developments we are going to
present in this paper.

The idea is to build up an unifying theory for all the plateaux with even
denominator starting from the bosonic Laughlin filling $\nu =1/pm+2$, which
is described by a CFT with $c=1$, in terms of a scalar chiral field
compactified on a circle with radius $R^{2}=1/\nu =pm+2$ (or the dual $%
R^{2}=4/pm+2$). Then the $U(1)$ current is given by $J(z)=i\partial _{z}Q(z)$%
, where $Q(z)$ is the compactified Fubini field with the standard mode
expansion (\ref{modes}). Let us notice that the informations about the
quantization of momentum and the winding numbers are stored in the lattice
geometry induced by the QHE quantization (see Ref. \cite{cgm2} for details);
in other words the QHE physics fixes the compactification radius. The
corresponding primary fields are expressed in terms of the vertex operators $%
U^{\alpha }(z)=:e^{i\alpha Q(z)}:$ with $\alpha ^{2}=1,...,2+pm$ and
conformal dimension $h=\frac{\alpha ^{2}}{2}$. Also here, as for Jain
fillings, starting with this set of fields and using the $m$-reduction
procedure, we get the image of the twisted sector of a $c=m$ orbifold CFT,
which describes the lowest Landau level dynamics.

Let us now concentrate on the special $m=2$ case, which describes a system
consisting of two parallel layers of $2D$ electrons gas in a strong
perpendicular magnetic field. The filling factor $\nu ^{(a)}=\frac{1}{2p+2}$
is the same for the two $a=1$, $2$ layers while the total filling is $\nu
=\nu ^{(1)}+\nu ^{(2)}=\frac{1}{p+1}$. For $p=0$ ($p=1$) it describes the
bosonic $220$ (fermionic $331$) Halperin (H) state \cite{Halperin}.

The CFT description for such a system can be given in terms of two
compactified chiral bosons $Q^{(a)}$ with central charge $c=2$. In order to
construct the fields $Q^{(a)}$ for the TM, the starting point is the bosonic
filling $\nu =1/2(p+1)$, described by a CFT with $c=1$ in terms of a scalar
chiral field $Q$ compactified on a circle with radius $R^{2}=1/\nu =2(p+1)$
(or its dual $R^{2}=2/(p+1)$), see Eq. (\ref{modes}). The $m$-reduction
procedure generates a daughter theory which is a $c=2$ orbifold. Its primary
fields content can be expressed in terms of a $Z_{2}$-invariant scalar field
$X(z)$, given by
\begin{equation}
X(z)=\frac{1}{2}\left( Q^{(1)}(z)+Q^{(2)}(-z)\right) ,  \label{X1}
\end{equation}
describing the electrically charged sector of the new filling, and a twisted
field
\begin{equation}
\phi (z)=\frac{1}{2}\left( Q^{(1)}(z)-Q^{(2)}(-z)\right) ,  \label{phi1}
\end{equation}
which satisfies the twisted boundary conditions $\phi (e^{i\pi }z)=-\phi (z)$
and describes the neutral sector \cite{cgm2}. Such TBC signal the presence
of a localized topological defect which couples the $2$ edges in such a way
to get a crossing, as sketched in Fig. 3.

\begin{figure}[ht]
\centering\includegraphics*[width=0.8\linewidth]{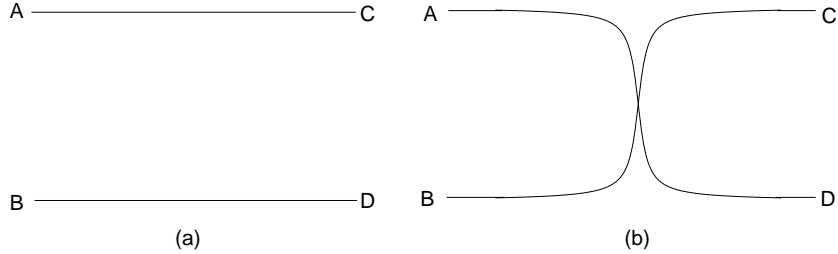}
\caption{The bilayer system, (a) without the topological defect (PBC), (b)
with the topological defect (TBC).}
\label{figura2}
\end{figure}

The chiral fields $Q^{(a)}$, defined on a single layer $a=1$, $2$, due to
the boundary conditions imposed upon them by the orbifold construction, can
be thought of as components of a unique ``boson''\ defined on a double
covering of the disc (layer) ($z_{i}^{(1)}=-z_{i}^{(2)}=z_{i}$). As a
consequence the two layers system becomes equivalent to one-layer QHF (in
contrast with the Halperin model in which they appear independent) and the $%
X $ and $\phi $ fields defined in Eqs. (\ref{X1}) and (\ref{phi1})
diagonalize the interlayer interaction. In particular the $X$ field carries
the total charge with velocity $v_{X}$, while $\phi $ carries the charge
difference of the two edges with velocity $v_{\phi }$, i.e. no charge, being
the number of electrons the same for each layer (balanced system).

The primary fields are the composite operators $V(z)=\mathcal{U}_{X}(z)\psi
(z)$, where $\mathcal{U}_{X}(z)=\frac{1}{\sqrt{z}}:e^{i\alpha X(z)}:$ are
the vertices of the charged sector with $\alpha ^{2}=2(p+1)$. Furthermore
the highest weight states of the neutral sector can be classified in terms
of two kinds of chiral operators, $\psi (z)\left( \bar{\psi}(z)\right) =%
\frac{1}{2\sqrt{z}}\left( e^{i\alpha {\cdot }\phi (z)}\pm e^{i\alpha {\cdot }%
\phi (-z)}\right) $, which, in a fermionic language, correspond to $c=1/2$
Majorana fermions with periodic (Ramond) or anti-periodic (Neveu-Schwarz)
boundary conditions \cite{cgm4}. As a consequence this theory decomposes
into a tensor product of two CFTs, a twisted invariant one with $c=3/2$,
realized by the charged boson $X(z)$ and the Ramond Majorana fermion,\ which
is coupled to the charged sector, while the second one has $c=1/2$ and is
realized in terms of the Neveu-Schwarz Majorana fermion. The two Majorana
fermions just defined are inequivalent, due to the breaking of the symmetry
which exchanges them, and that results in a non-Abelian statistics. Such a
factorization is much more evident in the construction of the modular
invariant partition function, as we briefly recall in the Appendix \cite
{cgm4}. The bosonized energy-momentum tensor of the $Z_{2}$ twist invariant
theory develops a cosine term in its neutral sector which is described by
the Ramond fields:
\begin{equation}
T_{\psi _{s}}(z)=-\frac{1}{4}(\partial \phi )^{2}-\frac{1}{16z^{2}}\cos (2%
\sqrt{2}\phi ).
\end{equation}
It is a clear signature of a tunneling phenomenon which selects out a new
stable vacuum,\ the $c=3/2$ one. If we refer to the bilayer system, we can
reduce the spacing between the layers so that the two species of electrons
which live on them become indistinguishable: in such a case the tunneling
amplitude gets large enough to make the H states flow to the Moore-Read (MR)
states \cite{MR}. In the limit of strong tunneling the velocity of one
Majorana becomes zero and the theory reduces to the $c=3/2$ CFT. Let us also
point out that $m$-ality in the neutral sector is coupled to the charged one
exactly, according to the physical request of locality of the electrons with
respect to the edge excitations. Indeed our projection, when applied to a
local field, automatically couples the discrete $Z_{m}$ charge of $U(1)$
with the neutral sector in order to give rise to a single valued composite
field.

Now let us give an interpretation of the existence of these sectors in terms
of conformal invariant boundary conditions which are due to the scattering
of the particles on localized impurities \cite{noi1}\cite{noi2}\cite{noi5}.
The H sector describes a pure QHF phase in which \ no impurities are present
and the two layers edges are not connected (see Fig. 3(a)). In realistic
samples however this is not the case and the deviations from the Halperin
state may be regarded as due to the presence of localized impurities. These
effects can be accounted for by allowing for more general boundary
conditions just as the ones provided by our TM. In fact an impurity located
at a given point on the edge induces twisted boundary conditions for the
boson $\phi $ and, as a consequence, a current can flow between the layers.
Then a coherent superposition of interlayer interactions could drive the
bilayer to a more symmetric phase in which the two layers are
indistinguishable due to the presence of a one electron tunneling effect
along the edge.

The primary fields content of the theory just introduced on the torus
topology will be given in the Appendix.

\section{m-reduction, noncommutativity and Morita equivalence}

In this Section we show how the issue of noncommutativity enters our CFT
description for QHF by fully exploiting the notion of Morita equivalence on
noncommutative tori; as we will see, it will be crucial to choose rational
values of the noncommutativity parameter $\theta $. That allows us to build
up a general isomorphism between NCFTs and CFTs on the ordinary space. We
will make an explicit reference to the $m$-reduced theory describing a QHF
at Jain and paired states fillings, recalled in Section 2. We obtain two
main results: i) from a theoretical perspective, a new characterization of
the $m$-reduction procedure is derived, as the image in the ordinary space
of Morita duality; ii) from a more applicative perspective, a new
relationship emerges between noncommutativity and QHF physics. Furthermore
the noncommutative torus Lie algebra is naturally realized, within the QHF
context, in terms of GMT.

The Morita equivalence \cite{morita}\cite{scw2} is an isomorphism between
noncommutative algebras that conserves all the modules and their associated
structures. Let us consider an $U(N)$ NCFT defined on the noncommutative
torus $\mathrm{T}_{\theta }^{2}$ and, for simplicity, of radii $R$. The
coordinates satisfy the commutation rule $[x_{1},x_{2}]=i\theta $ \cite
{ncft1}. In such a simple case the Morita duality is represented by the
following $SL(2,Z)$ action on the parameters:
\begin{equation}
\theta ^{^{\prime }}=\frac{a\theta +b}{c\theta +d};\text{ \ \ \ \ \ \ \ \ }%
R^{^{\prime }}=\left| c\theta +d\right| R,  \label{morita1}
\end{equation}
where $a,b,c,d$ are integers and $ad-bc=1$.

For rational values of the non commutativity parameter, $\theta =-\frac{b}{a}
$, so that $c\theta +d=\frac{1}{a}$, the Morita transformation (\ref{morita1}%
) sends the NCFT to an ordinary one with $\theta ^{^{\prime }}=0$ and
different radius $R^{^{\prime }}=\frac{R}{a}$, involving in particular a
rescaling of the rank of the gauge group \cite{alvarez1}\cite{moreno1}\cite
{moreno2}\cite{schiappa}. Indeed the dual theory is a twisted $U(N^{^{\prime
}})$ theory with $N^{^{\prime }}=aN$. The classes of $\theta ^{^{\prime }}=0$
theories are parametrized by an integer $m$, so that for any $m$ there is a
finite number of Abelian theories which are related by a subset of the
transformations given in Eq. (\ref{morita1}). This is a crucial remark as we
will show in the following, for CFT theories describing QHF at Jain as well
as paired states fillings, respectively.

\subsection{Jain fillings}

Let us show in detail how Morita duality works for Jain fillings. Indeed the
$m$-reduction technique applied to the QHF at Jain fillings ($\nu =\frac{m}{%
2pm+1}$) can be viewed as the image of the Morita map (characterized by $%
a=2p(m-1)+1$, $b=2p$, $c=m-1$, $d=1$) between the two NCFTs with $\theta =1$
and $\theta =2p+\frac{1}{m}$ respectively and corresponds to the Morita map
in the ordinary space. The $\theta =1$ theory is an $U\left( 1\right)
_{\theta =1}$ NCFT while the mother CFT is an ordinary $U\left( 1\right) $
theory; furthermore, when the $U\left( 1\right) _{\theta =2p+\frac{1}{m}}$
NCFT is considered, its Morita dual CFT has $U\left( m\right) $ symmetry. As
a consequence, the following correspondence Table between the NCFTs and the
ordinary CFTs is established:
\begin{equation}
\begin{array}{ccc}
& \text{Morita} &  \\
U\left( 1\right) _{\theta =1} & \rightarrow & U\left( 1\right) _{\theta =0}
\\
& \left( a=1,b=-1,c=0,d=1\right) &  \\
\text{Morita}\downarrow \left( a,b,c,d\right) &  & m-\text{reduction}%
\downarrow \\
& \text{Morita} &  \\
U\left( 1\right) _{\theta =2p+\frac{1}{m}} & \rightarrow & U\left( m\right)
_{\theta =0} \\
& \left( a=m,b=-2pm-1,c=1-m,d=2p\left( m-1\right) +1\right) &
\end{array}
\label{morita2}
\end{equation}

For more general commutativity parameters $\theta =\frac{q}{m}$ such a
correspondence can be easily extended. Indeed the action of the $m$%
-reduction procedure on the number $q$ doesn't change the central charge of
the CFT under study but modifies the compactification radius of the charged
sector \cite{cgm1}\cite{cgm3}. Nevertheless here we are interested to the
action of the Morita map on the denominator of the parameter $\theta $ which
has interesting consequences on noncommutativity, so in the following we
will concentrate on such an issue.

In order to show that the $m$-reduction technique applied to the QHF at Jain
fillings is the image of the Morita map between the two NCFTs with $\theta
=1 $ and $\theta =2p+\frac{1}{m}$ respectively and corresponds to the Morita
map in the ordinary space it is enough to show how the twisted boundary
conditions on the neutral fields of the $m$-reduced theory (see Section 2)
arise as a consequence of the noncommutative nature of the $U\left( 1\right)
_{\theta =2p+\frac{1}{m}}$ NCFT.

In order to carry out this program let us recall that an associative algebra
of smooth functions over the noncommutative two-torus $\mathrm{T}_{\theta
}^{2}$ can be realized through the Moyal product ($\left[ x_{1},x_{2}\right]
=i\theta $):
\begin{equation}
f\left( x\right) \ast g\left( x\right) =\left. \exp \left( \frac{i\theta }{2}%
\left( \partial _{x_{1}}\partial _{y_{2}}-\partial _{x_{2}}\partial
_{y_{1}}\right) \right) f\left( x\right) g\left( y\right) \right| _{y=x}.
\label{moyal1}
\end{equation}
It is convenient to decompose the elements of the algebra, i. e. the fields,
in their Fourier components. However a general field operator $\Phi $
defined on a torus can have different boundary conditions associated to any
of the compact directions. For the torus we have four different
possibilities:
\begin{equation}
\begin{array}{cc}
\Phi \left( x_{1}+R,x_{2}\right) =e^{2\pi i\alpha _{1}}\Phi \left(
x_{1},x_{2}\right) , & \Phi \left( x_{1},x_{2}+R\right) =e^{2\pi i\alpha
_{2}}\Phi \left( x_{1},x_{2}\right) ,
\end{array}
\label{bc1}
\end{equation}
where $\alpha _{1}$ and $\alpha _{2}$ are the boundary parameters. The
Fourier expansion of the general field operator $\Phi _{\overrightarrow{%
\alpha }}$ with boundary conditions $\overrightarrow{\alpha }=\left( \alpha
_{1},\alpha _{2}\right) $ takes the form:
\begin{equation}
\Phi _{\overrightarrow{\alpha }}=\sum_{\overrightarrow{n}}\Phi ^{%
\overrightarrow{n}}U_{\overrightarrow{n}+\overrightarrow{\alpha }}
\label{fexp1}
\end{equation}
where we define the generators as
\begin{equation}
U_{\overrightarrow{n}}\equiv \exp \left( 2\pi i\frac{\overrightarrow{n}\cdot
\overrightarrow{x}}{R}\right) .  \label{fexp2}
\end{equation}
They give rise to the following commutator:
\begin{equation}
\left[ U_{\overrightarrow{n}+\overrightarrow{\alpha }},U_{\overrightarrow{%
n^{\prime }}+\overrightarrow{\alpha ^{\prime }}}\right] =-2i\sin \left(
\frac{2\pi ^{2}\theta }{R^{2}}\left( \overrightarrow{n}+\overrightarrow{%
\alpha }\right) \wedge \left( \overrightarrow{n^{\prime }}+\overrightarrow{%
\alpha ^{\prime }}\right) \right) U_{\overrightarrow{n}+\overrightarrow{%
n^{\prime }}+\overrightarrow{\alpha }+\overrightarrow{\alpha ^{\prime }}},
\label{fexp3}
\end{equation}
where $\overrightarrow{p}\wedge \overrightarrow{q}=\varepsilon
_{ij}p_{i}q_{j}$.

When the noncommutativity parameter $\theta $ takes the rational value $%
\theta =\frac{2q}{m}\frac{R^{2}}{2\pi }$, being $q$ and $m$ relatively prime
integers, the infinite-dimensional algebra generated by the $U_{%
\overrightarrow{n}+\overrightarrow{\alpha }}$ breaks up into equivalence
classes of finite dimensional subspaces. Indeed the elements $U_{m%
\overrightarrow{n}}$ generate the center of the algebra and that makes
possible for the momenta the following decomposition:
\begin{equation}
\overrightarrow{n^{\prime }}+\overrightarrow{\alpha }=m\overrightarrow{n}+%
\overrightarrow{n},\text{ \ \ \ \ }0\leq n_{1},n_{2}\leq m-1.  \label{fexp4}
\end{equation}
The whole algebra splits into equivalence classes classified by all the
possible values of $m\overrightarrow{n}$, each class being a subalgebra
generated by the $m^{2}$ functions $U_{\overrightarrow{n}+\overrightarrow{%
\alpha }}$ which satisfy the relations
\begin{equation}
\left[ U_{\overrightarrow{n}+\overrightarrow{\alpha }},U_{\overrightarrow{%
n^{\prime }}+\overrightarrow{\alpha ^{\prime }}}\right] =-2i\sin \left(
\frac{\pi q}{m}\left( \overrightarrow{n}+\overrightarrow{\alpha }\right)
\wedge \left( \overrightarrow{n^{\prime }}+\overrightarrow{\alpha ^{\prime }}%
\right) \right) U_{\overrightarrow{n}+\overrightarrow{n^{\prime }}+%
\overrightarrow{\alpha }+\overrightarrow{\alpha ^{\prime }}}.  \label{fexp5}
\end{equation}
The algebra (\ref{fexp5}) is isomorphic to the (complexification of the) $%
U\left( m\right) $ algebra, whose general $m$-dimensional representation can
be constructed by means of the following ''shift'' and ''clock'' matrices
\cite{matrix2}\cite{matrix3}:
\begin{equation}
\mathrm{Q}=\left(
\begin{array}{cccc}
1 &  &  &  \\
& \varepsilon  &  &  \\
&  & \ddots  &  \\
&  &  & \varepsilon ^{m-1}
\end{array}
\right) ,\text{ \ \ \ \ \ \ }\mathrm{P}=\left(
\begin{array}{cccc}
0 & 1 &  & 0 \\
& \cdots  &  &  \\
&  & \vdots  & 1 \\
1 &  &  & 0
\end{array}
\right) ,  \label{fexp6}
\end{equation}
being $\varepsilon =\exp (\frac{2\pi iq}{m})$. So the matrices $J_{%
\overrightarrow{n}}=\varepsilon ^{n_{1}n_{2}}\mathrm{Q}^{n_{1}}\mathrm{P}%
^{n_{2}}$, $n_{1},n_{2}=0,...,m-1$, generate an algebra isomorphic to (\ref
{fexp5}):
\begin{equation}
\left[ J_{\overrightarrow{n}},J_{\overrightarrow{n^{\prime }}}\right]
=-2i\sin \left( \pi \frac{q}{m}\overrightarrow{n}\wedge \overrightarrow{%
n^{\prime }}\right) J_{\overrightarrow{n}+\overrightarrow{n^{\prime }}}.
\label{fexp7}
\end{equation}
Thus the following Morita mapping has been realized between the Fourier
modes defined on a noncommutative torus and functions taking values on $%
U\left( m\right) $ but defined on a commutative space:
\begin{equation}
\exp \left( 2\pi i\frac{\left( \overrightarrow{n}+\overrightarrow{\alpha }%
\right) \cdot \widehat{\overrightarrow{x}}}{R}\right) \longleftrightarrow
\exp \left( 2\pi i\frac{\left( \overrightarrow{n}+\overrightarrow{\alpha }%
\right) \cdot \overrightarrow{x}}{R}\right) J_{\overrightarrow{n}+%
\overrightarrow{\alpha }}.  \label{fexp8}
\end{equation}
As a consequence a mapping between the fields $\Phi _{\overrightarrow{\alpha
}}$ is generated as follows. Let us focus, for simplicity, on the case $q=1$
which leads for the momenta to the decomposition $\overrightarrow{n}=m%
\overrightarrow{n}+\overrightarrow{j}$, with \ \ \ \ $0\leq j_{1},j_{2}\leq m
$. The general field operator $\Phi _{\overrightarrow{\alpha }}$ on the
noncommutative torus $\mathrm{T}_{\theta }^{2}$ with boundary conditions $%
\overrightarrow{\alpha }$ can be written in the form:
\begin{equation}
\Phi _{\overrightarrow{\alpha }}=\sum_{\overrightarrow{n}}\exp \left( 2\pi im%
\frac{\overrightarrow{n}\cdot \overrightarrow{x}}{R}\right) \sum_{%
\overrightarrow{j}=0}^{m^{\prime }-1}\Phi ^{\overrightarrow{n},%
\overrightarrow{j}}U_{\overrightarrow{j}+\overrightarrow{\alpha }}.
\label{fexp9}
\end{equation}
By using Eq. (\ref{fexp8}) we obtain the Morita correspondence between
fields as:
\begin{equation}
\Phi _{\overrightarrow{\alpha }}\longleftrightarrow \Phi =\sum_{%
\overrightarrow{j}=0}^{m^{\prime }-1}\chi ^{\left( \overrightarrow{j}\right)
}J_{\overrightarrow{j}+\overrightarrow{\alpha }},  \label{fexp10}
\end{equation}
where we have defined:
\begin{equation}
\chi ^{\left( \overrightarrow{j}\right) }=\exp \left( 2\pi i\frac{\left(
\overrightarrow{j}+\overrightarrow{\alpha }\right) \cdot \overrightarrow{x}}{%
R}\right) \sum_{\overrightarrow{n}}\Phi ^{\overrightarrow{n},\overrightarrow{%
j}}\exp \left( 2\pi im\frac{\overrightarrow{n}\cdot \overrightarrow{x}}{R}%
\right) .  \label{fexp11}
\end{equation}
The field $\Phi $ is defined on the dual torus with radius $R^{\prime }=%
\frac{R}{m^{\prime }}$ and satisfies the boundary conditions:
\begin{equation}
\begin{array}{cc}
\Phi \left( \theta +R^{\prime },x_{2}\right) =\Omega _{1}^{+}\cdot \Phi
\left( \theta ,x_{2}\right) \cdot \Omega _{1}, & \Phi \left( \theta
,x_{2}+R^{\prime }\right) =\Omega _{2}^{+}\cdot \Phi \left( \theta
,x_{2}\right) \cdot \Omega _{2},
\end{array}
\label{fexp12}
\end{equation}
with
\begin{equation}
\Omega _{1}=\mathrm{P}^{b},\text{ \ \ \ \ \ \ }\Omega _{2}=\mathrm{Q}^{1/q},
\label{fexp13}
\end{equation}
where $b$ is an integer satisfying $am-bq=1$. While the field components $%
\chi ^{\left( \overrightarrow{j}\right) }$ satisfy the following twisted
boundary conditions:
\begin{equation}
\begin{array}{c}
\chi ^{\left( \overrightarrow{j}\right) }\left( \theta +R^{\prime
},x_{2}\right) =e^{2\pi i\left( j_{1}+\alpha _{1}\right) /m}\chi ^{\left(
\overrightarrow{j}\right) }\left( \theta ,x_{2}\right)  \\
\chi ^{\left( \overrightarrow{j}\right) }\left( \theta ,x_{2}+R^{\prime
}\right) =e^{2\pi i\left( j_{2}+\alpha _{2}\right) /m}\chi ^{\left(
\overrightarrow{j}\right) }\left( \theta ,x_{2}\right)
\end{array}
,  \label{fexp14}
\end{equation}
that is
\begin{equation}
\left( \frac{j_{1}+\alpha _{1}}{m},\frac{j_{2}+\alpha _{2}}{m}\right) ,\text{
\ \ \ \ }j_{1}=0,...,m-1,\text{ \ \ \ \ }j_{2}=0,...,m-1.  \label{fexp15}
\end{equation}
Let us observe that $\overrightarrow{j}=\left( 0,0\right) $ is the trace
degree of freedom which can be identified with the $U(1)$ component of the
matrix valued field or the charged component within the $m$-reduced theory
of the QHF at Jain fillings introduced in Section 2. We infer that only the
integer part of $\frac{n_{i}}{m}$ should really be thought of as the
momentum. The commutative torus is smaller by a factor $m\times m$ than the
noncommutative one; in fact, upon this rescaling, also the ''density of
degrees of freedom'' is kept constant as now we are dealing with $m\times m$
matrices instead of scalars.

Summarizing, when the parameter $\theta $ is rational we recover the whole
structure of the noncommutative torus and recognize the twisted boundary
conditions which characterize the neutral fields (\ref{phi}) of the $m$%
-reduced theory as the consequence of the Morita mapping of the starting
NCFT ($U\left( 1\right) _{\theta =2p+\frac{1}{m}}$ in our case) on the
ordinary commutative space. Indeed $\chi ^{\left( 0,0\right) }$ corresponds
to the charged $X$ field while the twisted fields $\chi ^{\left(
\overrightarrow{j}\right) }$ with $\overrightarrow{j}\neq \left( 0,0\right) $
should be identified with the neutral ones (\ref{phi}). Therefore the $m$%
-reduction technique can be viewed as a realization of the Morita mapping
between NCFTs and CFTs on the ordinary space, as sketched in the Table (\ref
{morita2}).

Let us now complete the proof by introducing generalized magnetic
translations which realize the noncommutative torus Lie algebra defined in
Eq. (\ref{fexp5}). In order to define GMT let us point out that, in our TM
model for the QHF (see Section 2 and Refs. \cite{cgm1},\cite{cgm3},\cite
{cgm2},\cite{cgm4}), the primary fields (and then the corresponding
characters within the torus topology) appear as composite field operators
which factorize in a charged as well as a neutral part. Further they are
also coupled by the discrete symmetry group $Z_{m}$. This decomposition must
hold for magnetic translations as well, so we need to generalize them in
such a way that they will appear as operators with two factors, acting on
the charged and on the neutral sector respectively. The presence of the
transverse magnetic field $B$ reduces the torus to a noncommutative one and
the flux quantization induces rational values of the noncommutativity
parameter $\theta $.

Let us also recall that the incompressibility of the quantum Hall fluid
naturally leads to a $W_{1+\infty }$ dynamical symmetry \cite{BS, ctz}.
Indeed, if one considers a droplet of a quantum Hall fluid, it is evident
that the only possible area preserving deformations of this droplet are the
waves at the boundary of the droplet, which describe the deformations of its
shape, the so called edge excitations. These can be well described by the
infinite generators $W_{m}^{n+1}$ of $W_{1+\infty }$ of conformal spin $%
(n+1) $, which are characterized by a mode index $m\in Z$ and satisfy the
algebra:
\begin{equation}
\left[ W_{m}^{n+1},W_{m^{\prime }}^{n^{\prime }+1}\right] =(n^{\prime
}m-nm^{\prime })W_{m+m^{\prime }}^{n+n^{\prime }}+q(n,n^{\prime
},m,m^{\prime })W_{m+m^{\prime }}^{n+n^{\prime }-2}+...+d(n,m)c\,\delta
^{n,n^{\prime }}\delta _{m+m^{\prime }=0},  \label{walgebras}
\end{equation}
where the structure constants $q$ and $d$ are polynomials of their
arguments, $c$ is the central charge, and dots denote a finite number of
similar terms involving the operators $W_{m+m^{\prime }}^{n+n^{\prime }-2l}$
\cite{BS, ctz}. Such an algebra contains an Abelian $\widehat{U}(1)$ current
for $n=0$ and a Virasoro algebra for $n=1$ with central charge $c$. It
encodes the local properties which are imposed by the incompressibility
constraint and realizes the allowed edge excitations \cite{ctz1}.
Nevertheless algebraic properties do not include topological properties
which are also a consequence of incompressibility. In order to take into
account the topological properties we have to resort to finite magnetic
translations which encode the large scale behavior of the QHF.

Let us consider a magnetic translation of step $(n=n_{1}+in_{2},\overline{n}%
=n_{1}-in_{2})$ on a sample with coordinates $x_{1},x_{2}$ and define the
corresponding generators $T^{n,\overline{n}}$ as:
\begin{equation}
T^{n,\overline{n}}=e^{-\frac{B}{4}n\overline{n}}e^{\frac{1}{2}nb^{+}}e^{%
\frac{1}{2}\overline{n}b},  \label{tr1}
\end{equation}
where $b^{+}=i\partial _{\overline{\omega }}-i\frac{B}{2}\omega $, $%
b=i\partial _{\omega }+i\frac{B}{2}\overline{\omega }$, $\omega $ is a
complex coordinate ($\overline{\omega }$ being its conjugate) and $B$ is the
transverse magnetic field. They satisfy the relevant property:
\begin{equation}
T^{n,\overline{n}}T^{m,\overline{m}}=q^{-\frac{n\times m}{4}}T^{n+m,}{}^{%
\overline{n}+\overline{m}},  \label{tr2}
\end{equation}
where $q$ is a root of unity.

Furthermore, it can be easily shown that they admit the following expansion
in terms of the generators $W_{k-1}^{l-1}$ of the $W_{1+\infty }$ algebra:
\begin{equation}
T^{n,\overline{n}}=e^{-\frac{B}{4}n\overline{n}}\sum_{k,l=0}^{\infty }\left(
-\right) ^{l}\frac{n^{k}}{2^{k}n}\frac{\overline{n}^{l}}{2^{l}\overline{n}}%
W_{k-1}^{l-1},  \label{tr3}
\end{equation}
where now the local $W_{1+\infty }$ symmetry and the global topological
properties are much more evident because the coefficients in the above
series depend on the topology of the sample.

Within our $m$-reduced theory for a QHF at Jain fillings \cite{cgm1}\cite
{cgm3}, introduced in Section 2, it can be shown that also magnetic
translations of step $(n,\overline{n})$ decompose into equivalence classes
and can be factorized into a group, with generators $T_{C}^{n,\overline{n}}$%
, which acts only on the charged sector as well as a group, with generators $%
T_{S}^{j_{i},\overline{j}}$, acting only on the neutral sector. The presence
of the transverse magnetic field $B$ reduces the torus to a noncommutative
one and the flux quantization induces rational values of the
noncommutativity parameter $\theta $. As a consequence the neutral magnetic
translations realize a projective representation of the $su\left( m\right) $
algebra generated by the elementary translations:
\begin{equation}
J_{a,b}=e^{-2\pi i\frac{ab}{m}}T_{S}^{a,0}T_{S}^{0,b};\text{ \ \ \ \ \ \ }%
a,b=1,...,m,  \label{tr4}
\end{equation}
which satisfy the commutation relations:
\begin{equation}
\left[ J_{a,b},J_{\alpha ,\beta }\right] =-2i\sin \left( \frac{2\pi }{m}%
\left( a\beta -b\alpha \right) \right) J_{a+\alpha ,b+\beta }.  \label{tr5}
\end{equation}
The GMT operators above defined (see Eqs. (\ref{tr1}) and (\ref{tr4})) are a
realization of the operators introduced in Eq. (\ref{fexp2}) and the algebra
defined by Eq. (\ref{tr5}) is isomorphic to the noncommutative torus Lie
algebra given in Eqs. (\ref{fexp5}) and (\ref{fexp7}). Such operators
generate the residual symmetry of the $m$-reduced CFT which is Morita
equivalent to the NCFT with rational non commutativity parameter $\theta =2p+%
\frac{1}{m}$.

\subsection{Paired states fillings}

In order to show how Morita duality works also for QHF at paired states
fillings, let us proceed as in the previous Subsection.

Let us put our $m$-reduced theory on a two-torus and consider its
noncommutative counterpart $\mathrm{T}_{\theta }^{2}$, where $\theta $ is
the noncommutativity parameter. The $m$-reduction technique applied to the
QHF at paired states fillings ($\nu =\frac{m}{pm+2}$, $p$ even) can be
viewed as the image of the Morita map \cite{morita}\cite{scw1}\cite{scw2}
(characterized by $a=\frac{p}{2}\left( m-1\right) +1$, $b=\frac{p}{2}$, $%
c=m-1$, $d=1$) between the two NCFTs with $\theta =1$ and $\theta =\frac{p}{2%
}+\frac{1}{m}$ ($\theta =\nu _{0}/\nu ,$ being $\nu _{0}=1/2$ the filling of
the starting theory) respectively, and corresponds to the Morita map in the
ordinary space. The $\theta =1$ theory is an $U\left( 1\right) _{\theta =1}$
NCFT while the mother CFT is an ordinary $U\left( 1\right) $ theory;
furthermore, when the $U\left( 1\right) _{\theta =\frac{p}{2}+\frac{1}{m}}$
NCFT is considered, its Morita dual CFT has $U\left( m\right) $ symmetry. As
a consequence, the following correspondence Table between the NCFTs and the
ordinary CFTs is established \cite{AV2}:
\begin{equation}
\begin{array}{ccc}
& \text{Morita} &  \\
U\left( 1\right) _{\theta =1} & \rightarrow  & U\left( 1\right) _{\theta =0}
\\
& \left( a=1,b=-1,c=0,d=1\right)  &  \\
\text{Morita}\downarrow \left( a,b,c,d\right)  &  & m-\text{reduction}%
\downarrow  \\
& \text{Morita} &  \\
U\left( 1\right) _{\theta =\frac{p}{2}+\frac{1}{m}} & \rightarrow  & U\left(
m\right) _{\theta =0} \\
& \left( a=m,b=-\frac{pm}{2}-1,c=1-m,d=\frac{p}{2}\left( m-1\right)
+1\right)  &
\end{array}
\label{moritatb2}
\end{equation}
Let us notice that theories which differ by an integer in the
noncommutativity parameter are not identical because they differ from the
point of view of the CFT. In fact, the Morita map acts on more than one
parameter of the theory. For instance, the compactification radius of the
charged component is renormalized to $R_{X}^{2}=p+\frac{2}{m}$,\ that gives
rise to different CFTs by varying $p$ values. Moreover the action of the $m$%
-reduction procedure on the number $p$ doesn't change the central charge of
the CFT under study but modifies the spectrum of the charged sector \cite
{cgm2}\cite{cgm4}. Furthermore the twisted boundary conditions on the
neutral fields of the $m$-reduced theory, Eq. (\ref{phi1})), arise as a
consequence of the noncommutative nature of the $U\left( 1\right) _{\theta =%
\frac{p}{2}+\frac{1}{m}}$ NCFT.

Also here the key role in the proof of equivalence is played by the map on
the field $Q(z)$\ of Eq. (\ref{modes}) which, after the Morita action, is
defined on the noncommutative space $z\rightarrow z^{1/m}\equiv U_{0,1}$.
The noncommutative torus Lie algebra defined by the following commutation
rules:
\begin{equation}
\left[ U_{\overrightarrow{n}+\overrightarrow{j}},U_{\overrightarrow{%
n^{\prime }}+\overrightarrow{j^{\prime }}}\right] =-2i\sin \left( 2\pi
\theta \overrightarrow{j}\wedge \overrightarrow{j^{\prime }}\right) U_{%
\overrightarrow{n}+\overrightarrow{n^{\prime }}+\overrightarrow{j}+%
\overrightarrow{j^{\prime }}}.  \label{pai1}
\end{equation}
is realized in terms of the $m^{2}-1$ general operators:
\begin{equation}
\begin{array}{cc}
U_{j_{1},j_{2}}=\varepsilon ^{\frac{j_{1}j_{2}}{2}}z^{j_{1}}\varepsilon
^{j_{2}\widetilde{\sigma }}, &
\begin{array}{c}
j_{1},j_{2}=0,...,m-1 \\
\left( j_{1},j_{2}\right) \neq \left( 0,0\right)
\end{array}
\end{array}
,  \label{circles4}
\end{equation}
\bigskip where $\widetilde{\sigma }=iz\partial _{z}$. Via Morita duality, a
mapping between a general field operator $\mathbf{\Phi }$ defined on the
noncommutative torus \textrm{T}$_{\theta }^{2}$ and the field $\Phi $ living
on the dual commutative torus $\mathrm{T}_{\theta =0}^{2}$ is generated as
follows:
\begin{equation}
\mathbf{\Phi }=\sum_{\overrightarrow{n}}\exp \left( 2\pi im\frac{%
\overrightarrow{n}\cdot \widehat{\overrightarrow{x}}}{R}\right) \sum_{%
\overrightarrow{j}=0}^{m-1}\Phi ^{\overrightarrow{n},\overrightarrow{j}}U_{%
\overrightarrow{n}+\overrightarrow{j}}\longleftrightarrow \Phi =\sum_{%
\overrightarrow{j}=0}^{m-1}\chi ^{\left( \overrightarrow{j}\right) }J_{%
\overrightarrow{j}}.  \label{pai2}
\end{equation}
The new field $\Phi $ is defined on the dual torus with radius $R^{^{\prime
}}=\frac{R}{m}$ and satisfies the \textit{twist eaters} boundary conditions (%
\ref{fexp12}), while the field components $\chi ^{\left( \overrightarrow{j}%
\right) }$ satisfy twisted boundary conditions.

By using the above decomposition (Eq. (\ref{pai2})), where $\Phi \equiv Q(z)$%
, we identify the fields $X(z)$\ and $\phi ^{j}(z)$\ of the CFT defined on
the ordinary space. Indeed $\chi ^{\left( 0,0\right) }$ is the trace degree
of freedom which can be identified with the $U(1)$ component of the matrix
valued field or the charged $X$ field (\ref{X1}) within the $m$-reduced
theory of the QHF, while the twisted fields $\chi ^{\left( \overrightarrow{j}%
\right) }$ with $\overrightarrow{j}\neq \left( 0,0\right) $ should be
identified with the neutral ones (\ref{phi1}).

In conclusion, when the parameter $\theta $ is rational we recover the whole
structure of the noncommutative torus and recognize the twisted boundary
conditions which characterize the neutral fields (\ref{phi1}) of the $m$%
-reduced theory as the consequence of the Morita mapping of the starting
NCFT ($U\left( 1\right) _{\theta =\frac{p}{2}+\frac{1}{m}}$ in our case) on
the ordinary commutative space. In such a picture the GMT are a realization
of the noncommutative torus Lie algebra defined in Eq. (\ref{pai1}), as we
wiil show in detail in the following Sections by making explicit reference
to the bilayer case $m=2$.

Here we start only to outline the general structure of GMT, anticipating
some ideas which we will develop in detail in the following Sections. In
order to pursue this task, let us consider a general magnetic translation of
step $\left( n=n_{\upharpoonright }+in_{\downharpoonleft },\overline{n}%
=n_{\upharpoonright }-in_{\downharpoonleft }\right) $ on a sample with
coordinates $\left( x_{1},x_{2}\right) $ and denote with $T^{n,\overline{n}}$
the corresponding generators. Let us denote with $\upharpoonleft $ and $%
\downharpoonright $ the layer index because we are dealing with a bilayer
system. Within our TM for a QHF at paired states fillings \cite{cgm2}\cite
{cgm4} it is possible to show that such generators can be factorized into a
group which acts only on the charged sector as well as a group acting only
on the neutral sector \cite{AV3}. In this context the $\mathit{classical}$
magnetic translations group considered in the literature corresponds to the
TM charged sector. In order to study the action of a GMT on the torus and
clarify its interpretation in terms of noncommutative torus Lie algebra, Eq.
(\ref{pai1}), let us evaluate how the argument of the Theta functions in
which the conformal blocks are expressed gets modified. For a bilayer Hall
system a translation carried out on the layer $\upharpoonleft $ or $%
\downharpoonright $ produces a shift in the layer Theta argument $w_{i}$, $%
w_{i}\rightarrow w_{i}+\delta _{i}$, which can be conveniently expressed in
terms of the charged and neutral ones $w_{c(n)}=\frac{w_{\upharpoonleft }\pm
w_{\downharpoonright }}{2}$, and in this way we obtain the action on the
conformal blocks of the TM. Indeed, from the periodicity of the Theta
functions it is easy to show that the steps of the charged and neutral
translation can be parametrized by $\delta _{c}=\frac{2(p+1)l+2s+i}{2(p+1)}$
and $\delta _{n}=\pm l\pm \frac{i}{2}$ respectively, being $l=0,1$; $%
s=0,...,p$; and $i=0,1$. The layer exchange is realized by the
transformation $w_{n}\rightarrow -w_{n}$ but the TM is built in such a way
to correspond to the exactly balanced system in which $w_{n}=0$\ (modulo
periodicity) so that this operation can be obtained only by exchanging the
sign in $\delta _{n}$ (independently for $l$ and $i$). Because of the
factorization of the effective CFT at paired states fillings into two
sub-theories with $c=3/2$ and $c=1/2$, corresponding to the MR and Ising
model respectively (see Section 2, Appendix and Refs. \cite{cgm2}\cite{cgm4}%
), we infer that also GMT exhibit the same factorization \cite{AV3}. As a
consequence, conformal blocks of MR and Ising sectors are stable under the
transport of electrons and of the neutral Ising fermion.

In this way some aspects of the structure of GMT for the quantum hall
bilayer at paired states fillings become to emerge. In the following Section
we discuss in detail the bilayer physics and add new ingredients which will
help us to infer the general structure of GMT. Such a general structure will
be the subject of Section 5.

\section{Quantum Hall bilayer in the presence of a topological defect: the
role of boundary interactions}

The aim of this Section is to present in detail the physics of quantum Hall
bilayers in the presence of a topological defect localized somewhere on the
edge, in order to work out all the main features of our field theoretic
approach for a simple but non trivial system. We start by summarizing the
different possible boundary conditions of our CFT model for the quantum Hall
bilayer and then point out its equivalence with a system of two massless
scalar bosons with a magnetic boundary interaction at ``magic''\ points \cite
{noi1}\cite{noi2}\cite{noi5}. Then, starting from the boundary content of
our TM, we introduce the GMT action in a simple way, in terms of the
periodicity of the Jacobi theta functions which enter the boundary partition
functions $Z_{NB_{V}}\left( \delta ,V\right) $ \cite{noi2}, given in Eq. (%
\ref{ZUV1}). In particular we show how the defect interaction parameters $%
(V,\delta )$ change upon GMT. This behavior characterizes GMT as boundary
condition changing operators and allows us to infer their general structure,
which we present in Section 6.

Our TM theory is the continuum description of the quantum Hall bilayer under
study. Its key feature is the presence of two different boundary conditions
for the fields defined on the two layers:
\begin{equation}
\varphi _{L}^{\left( 1\right) }\left( x=0\right) =\pm \varphi _{R}^{\left(
2\right) }\left( x=0\right) ,  \label{blr}
\end{equation}
where the $+$ ($-$) sign identifies periodic (PBC) and twisted (TBC)
boundary conditions respectively, $L$ and $R$ staying for left and right
components. Indeed TBC are naturally satisfied by the twisted field $\phi
\left( z\right) $ of our TM (see Eq. (\ref{phi1})), which describes both the
left moving component $\varphi _{L}^{\left( 1\right) }$ and the right moving
one $\varphi _{R}^{\left( 2\right) }$ in a folded description of a system
with boundary. In the limit of strong coupling they account for the
interaction between a topological defect at the point $x=0$ (layers crossing
shown in Fig. 3) and the up and down edges of the bilayer system. When going
to the torus topology, the characters of the theory are in one to one
correspondence with the ground states and a doubling of the corresponding
degeneracy is expected, which can be seen at the level of the conformal
blocks (see Appendix). Indeed we get for the PBC case an untwisted sector, $%
P-P$ and $P-A$, described by the conformal blocks (\ref{vacuum1})-(\ref{ut3}%
), and for the TBC case a twisted sector, $A-P$ and $A-A$, described by the
conformal blocks (\ref{tw1})-(\ref{tw6}). Summarizing, the two layer edges
can be disconnected or connected in different ways, implying different
boundary conditions, which can be discussed referring to the characters with
the implicit relation to the different boundary states (BS) present in the
system (see Ref. \cite{noi1}). These BS should be associated to different
kinds of linear defects compatible with conformal invariance and their
relative stability can be established. The knowledge of the relative
stability of the different boundary states is crucial for the reconstruction
of the whole boundary renormalization group (RG) flow. Indeed different
boundary conditions correspond to different classes of boundary states, each
one characterized by a $g$-function \cite{Affleck2}, the $g$-function
decreases along the RG flow when going form the UV to the IR fixed point
\cite{noi1} and the generalized magnetic translations play the role of
boundary condition changing operators, as we will show in the following.

Let us now write down the action for our bilayer system in correspondence of
the different boundary conditions imposed upon it, i. e. PBC and TBC. In the
absence of an edge crossing (PBC case) the Hamiltonian of the bilayer system
is simply:
\begin{equation}
H=\frac{1}{2}\left[ \left( \Pi ^{\left( 1\right) }\right) ^{2}+\left( \Pi
^{\left( 2\right) }\right) ^{2}+\left( \partial _{x}Q^{(1)}\right)
^{2}+\left( \partial _{x}Q^{(2)}\right) ^{2}\right]  \label{hampbc}
\end{equation}
where $Q^{(1)}$ and $Q^{(2)}$ are the two boson fields generated by $2$%
-reduction and defined on the layers $1$ and $2$ respectively (see Section
2), while the presence of such a coupling (TBC case, see Fig. 3) introduces
a magnetic twist term of the kind:
\begin{equation}
H_{M}=\beta \left( Q^{(1)}\partial _{t}Q^{(2)}-Q^{(2)}\partial
_{t}Q^{(1)}\right) \delta \left( x\right) .  \label{magterm}
\end{equation}
Finally, in the presence of a localized defect (or a quantum point contact)
the Hamiltonian contains a boundary tunneling term such as:
\begin{equation}
H_{P}=-t_{P}\cos \left( Q^{(1)}-Q^{(2)}\right) \delta \left( x\right) ,
\label{pot1}
\end{equation}
which implements a locally applied gate voltage $V_{g}=t_{P}\delta \left(
x\right) $. Thus the full Hamiltonian can be written as \cite{noi2}\cite
{noi5}:
\begin{eqnarray}
H &=&\frac{1}{2}\left[ \left( \Pi ^{\left( 1\right) }\right) ^{2}+\left( \Pi
^{\left( 2\right) }\right) ^{2}+\left( \partial _{x}Q^{(1)}\right)
^{2}+\left( \partial _{x}Q^{(2)}\right) ^{2}\right] -t_{P}\cos \left(
Q^{(1)}-Q^{(2)}\right) \delta \left( x\right)  \nonumber \\
&&+\beta \left( Q^{(1)}\partial _{t}Q^{(2)}-Q^{(2)}\partial
_{t}Q^{(1)}\right) \delta \left( x\right) .
\end{eqnarray}
Introducing the charged and neutral fields $X$ and $\phi $ defined in Eqs. (%
\ref{X1}) and (\ref{phi1}) we clearly see that the boundary tunneling term
in the Hamiltonian is proportional to $\phi $ and the magnetic term produces
a twist on $\phi $.

Our bilayer system looks like very similar to a system of two massless
scalar fields $X$ and $Y$ \ in $1+1$ dimensions, which are free in the bulk
except for boundary interactions, which couple them. Its action is given by $%
S=S_{bulk}+S_{pot}+S_{mag}$ \cite{maldacena} where:
\begin{eqnarray}
S_{bulk} &=&\frac{\alpha }{4\pi }\int_{0}^{T}dt\int_{0}^{l}d\sigma \left(
\left( \partial _{\mu }X\right) ^{2}+\left( \partial _{\mu }Y\right)
^{2}\right) , \\
S_{pot} &=&\frac{V}{\pi }\int_{0}^{T}dt\left( \cos X\left( t,0\right) +\cos
Y\left( t,0\right) \right) ,  \label{SP} \\
S_{mag} &=&i\frac{\beta }{4\pi }\int_{0}^{T}dt\left( X\partial
_{t}Y-Y\partial _{t}X\right) _{\sigma =0}.  \label{SM}
\end{eqnarray}
Here $\alpha $ determines the strength of dissipation and is related to the
potential $V$ while $\beta $ is related to the strength of the magnetic
field $B$ orthogonal to the $X-Y$ plane, as $\beta =2\pi B$. The magnetic
term introduces a coupling between $X$ and $Y$ at the boundary while keeping
conformal invariance. Such a symmetry gets spoiled by the presence of the
interaction potential term except for the magic points $\left( \alpha ,\beta
\right) =\left( \frac{1}{n^{2}+1},\frac{n}{n^{2}+1}\right) ,n\in \mathbb{Z}$%
. For such parameters values the theory is conformal invariant for any
potential strength $V$. Furthermore, if $\alpha =\beta $ there is a complete
equivalence between our TM model for the bilayer system and the above
boundary CFT, as shown in Ref. \cite{noi2}. All the degrees of freedom of
such a system are expressed in terms of boundary states, which can be easily
constructed by considering the effect of the magnetic interaction term as
well as that of the potential term on the Neumann boundary state $|N>$. In
this way one obtains the generalized boundary state $|B_{V}>$ as:
\begin{equation}
|B_{V}>=\sec \left( \frac{\delta }{2}\right) e^{i\delta \mathcal{R}%
_{M}}e^{-H_{pot}\left( 2X_{L}^{^{\prime }}\right) -H_{pot}\left(
2Y_{L}^{^{\prime }}\right) }|N^{X^{^{\prime }}}>|N^{Y^{^{\prime }}}>,
\end{equation}
where the rotation operator $\mathcal{R}_{M}$ is given by
\begin{equation}
\mathcal{R}_{M}=(y_{L}^{0}p_{L}^{X}-x_{L}^{0}p_{L}^{Y})+\sum_{n>0}\frac{i}{n}%
\left( \alpha _{n}^{Y}\alpha _{-n}^{X}-\alpha _{-n}^{Y}\alpha
_{n}^{X}\right)   \label{rot1}
\end{equation}
and the rotation parameter $\delta $ is defined in terms of the parameters $%
\alpha ,\beta $ as $\tan \left( \frac{\delta }{2}\right) =\frac{\beta }{%
\alpha }$. Furthermore the rotated and rescaled coordinates $X^{^{\prime
}},Y^{^{\prime }}$ have been introduced as:
\begin{eqnarray}
X^{^{\prime }} &=&\cos \frac{\delta }{2}\left( \cos \frac{\delta }{2}X-\sin
\frac{\delta }{2}Y\right) ,  \nonumber \\
Y^{^{\prime }} &=&\cos \frac{\delta }{2}\left( \sin \frac{\delta }{2}X+\cos
\frac{\delta }{2}Y\right) .  \label{XY1}
\end{eqnarray}
Finally the boundary partition function $Z_{NB_{V}}$ can be computed as:
\begin{equation}
Z_{NB_{V}}=\sec \left( \frac{\delta }{2}\right) =<N|q^{L_{0}+\widetilde{L}%
_{0}}|B_{V}>  \label{part1}
\end{equation}
because, in the open string language, the rotation $\mathcal{R}_{M}$
introduces now twisted boundary conditions in the $\sigma $ direction. In
order to better clarify the equivalence of our twisted theory with the above
system of two massless scalar bosons with boundary interaction at ``magic''\
points it has been shown that the interlayer interaction is diagonalized by
the effective fields $X,\phi $ of Eqs. (\ref{X1}) and (\ref{phi1}), which
are related to the layers fields $Q^{(1)},Q^{(2)}$ just by the relation
given in Eq. (\ref{XY1}) for $\alpha =\beta $ \cite{noi2}. Indeed they can
be rewritten as:
\begin{eqnarray}
X(z) &=&\cos (\varphi /4)\left( \sin (\varphi /4)Q^{(1)}(z)+\cos (\varphi
/4)Q^{(2)}(z)\right) , \\
\phi (z) &=&\cos (\varphi /4)\left( \cos (\varphi /4)Q^{(1)}(z)-\sin
(\varphi /4)Q^{(2)}(z)\right) .
\end{eqnarray}
Such a transformation consists of a scale transformation plus a rotation;
for $\varphi =\pi $ the fields $X(z)$ and $\phi (z)$ of Eqs. (\ref{X1}) and (%
\ref{phi1}) are obtained and the transformations above coincide with the
transformations given in Eqs. (\ref{XY1}) for $\delta =\frac{\pi }{2}$. In
this context the boundary state $|B_{0}(\delta )>$ for the (untwisted)
twisted sector in the folded theory is obtained from the rotation $\mathcal{R%
}_{M}$ on the Neumann boundary state $|N(\theta )>$ when $\delta =0,\frac{%
\pi }{2}$ respectively and can be seen as due to a boundary magnetic term
according to \cite{maldacena}. Finally, by performing the rescaling $%
z\rightarrow z^{\frac{1}{2}}$, $a_{2n+l}\rightarrow \sqrt{2}a_{n+\frac{l}{2}}
$, $q\rightarrow \frac{q}{\sqrt{2}}$, for $\alpha =\beta $, we obtain the $%
X(z)$ and $\phi (z)$ fields in the standard form. As a result, the twisted
CFT can be conjectured to represent the correct CFT which describes
dissipative quantum mechanics of Ref. \cite{cgm5}.

Now we compute the boundary partition functions $Z_{NB_{V}}(\delta ,V)$ \cite
{noi2}, which express the boundary content of our theory, and show that they
are closed under the action of GMT. That will be proven by studying how the
defect interaction parameters $(V,\delta )$ change upon GMT. As a result we
find that the translations of electrons as well as anyons form a closed
algebra (the supersymmetric sine algebra (SSA) given in the next Section,
Eqs. (\ref{ssa1})-(\ref{ssa3})), where the parameters $V$ and $\delta $
remain unchanged modulo $m$. Indeed the stability algebra for a fixed point
identified by a fractional value of $(V,\delta )$ \ is given by this
subalgebra and a subset of the conformal blocks for the boundary partition
function $Z_{NB_{V}}(\delta ,V)$. Also the quasi-hole translations form a
closed algebra but the parameters $(V,\delta )$ will change together with
the corresponding boundary partition function. In this way the transition to
a different fixed point of the boundary flow is obtained or, in other words,
the switching between the untwisted and twisted vacua of our TM is realized.
As a result the role of GMT as boundary condition changing operators clearly
emerges.

As a first step, let us briefly recall the boundary content of our theory in
the simplest $m=2$, $p=0$ case, which corresponds to the first non-trivial
``magic''\ point $\alpha =\beta =\frac{1}{2}$ in Ref. \cite{maldacena}. The
action of the magnetic boundary term, (\ref{magterm}) or (\ref{SM}), on the
Neumann state $|N>$ is obtained by defining a pair of left-moving fermions
as:
\begin{equation}
\begin{array}{ccc}
\psi _{1}=c_{1}e^{\frac{i}{2}\left( Q^{\left( 2\right) }+Q^{\left( 1\right)
}\right) }=c_{1}e^{iX}, &  & \psi _{2}=c_{2}e^{-\frac{i}{2}\left( Q^{\left(
2\right) }-Q^{\left( 1\right) }\right) }=c_{2}e^{i\phi },
\end{array}
\end{equation}
where $c_{i}$, $i=1,2$ are cocycles necessary for the anticommutation. By
splitting the two Dirac fermions into real and imaginary parts, $\varphi
_{1}=\psi _{11}+i\psi _{12}$, $\varphi _{2}=\psi _{21}+i\psi _{22}$, we get
four left-moving Majorana fermions given by $\psi =\left( \psi _{11},\psi
_{12},\psi _{21},\psi _{22}\right) =\left( \cos X,\sin X,\cos \phi ,\sin
\phi \right) $ and a corresponding set of right-moving ones. In this new
language the magnetic boundary term acts only on the fourth Majorana fermion
as $\mathrm{R}_{M}=e^{2i\delta }$, where $\delta =0$ ($\delta =\frac{\pi }{2}
$) for the untwisted (twisted) sector of our theory, being its action the
identity for the other components, while the potential term acts on the
Majoranas as:
\begin{equation}
\mathrm{R}_{P}=\left(
\begin{array}{cccc}
\cos \left( 2V\right) & -\sin \left( 2V\right) & 0 & 0 \\
\sin \left( 2V\right) & \cos \left( 2V\right) & 0 & 0 \\
0 & 0 & \cos \left( 2V\right) & -\sin \left( 2V\right) \\
0 & 0 & \sin \left( 2V\right) & \cos \left( 2V\right)
\end{array}
\right) .
\end{equation}
So the overall rotation of the corresponding fermionic boundary states is $%
\mathrm{R}=\mathrm{R}_{M}\mathrm{R}_{P}$ and the partition function $Z_{AB}$%
, can be rewritten as:
\begin{equation}
Z_{NB_{V}}\left( \delta ,V\right) =\left\langle N\right| e^{-L\left( L_{0}+%
\bar{L}_{0}\right) }|B_{V}\left( \delta \right) >=\sqrt{2}\left( q\right)
^{-2/24}\prod_{n=1}^{\infty }\det \left( \mathrm{I}+q^{n-\frac{1}{2}}\mathrm{%
R}\right) ,
\end{equation}
where $|A>$ is the Neumann boundary state $|N>$, $|B>$ is the
magnetic-potential BS $|B_{V}>$, $q=e^{2i\pi \tau }$ and $\mathrm{I}$ is the
identity matrix. The final result is:
\begin{equation}
Z_{NB_{V}}\left( \delta ,V\right) =\sqrt{2}\left( \frac{\theta _{3}\left(
V|\tau \right) }{\eta \left( \tau \right) }\sqrt{\frac{\theta _{3}\left(
V|\tau \right) }{\eta \left( \tau \right) }}\right) \sqrt{\frac{\theta
_{3}\left( \delta +V|\tau \right) }{\eta \left( \tau \right) }},
\label{ZUV1}
\end{equation}
where $\delta =0$ ($\delta =\frac{\pi }{2}$) for the untwisted (twisted)
sector.

The value of the parameters $\delta $ and $V$ identifies a fixed point in
the boundary flow. Now, in order to compute the GMT action on these fixed
points and characterize GMT as boundary condition changing operators, let's
look at the transformation properties of the generalized Jacobi $\theta
_{i}\left( \omega |\tau \right) $ functions, $i=1,...,4$, for translations
along the cycles $A$ and $B$ of the two-torus. In particular, the
translations of interest for our study correspond to the following
transformations $\omega \rightarrow \omega +a+b\tau $ of the $\omega $
parameter with $a,b=\frac{1}{2},1$. The result is:

\begin{equation}
\begin{array}{cc}
\theta _{1}\left( 1|\tau \right) =\theta _{1}\left( 0|\tau \right) , &
\theta _{2}\left( 1|\tau \right) =-\theta _{2}\left( 0|\tau \right) \\
\theta _{3}\left( 1|\tau \right) =\theta _{3}\left( 0|\tau \right) , &
\theta _{4}\left( 1|\tau \right) =\theta _{4}\left( 0|\tau \right) \\
\theta _{1}\left( \tau |\tau \right) =-q^{-1}\theta _{1}\left( 0|\tau
\right) , & \theta _{2}\left( \tau |\tau \right) =q^{-1}\theta _{2}\left(
0|\tau \right) , \\
\theta _{3}\left( \tau |\tau \right) =q^{-\frac{1}{2}}\theta _{3}\left(
0|\tau \right) , & \theta _{4}\left( \tau |\tau \right) =-q^{-\frac{1}{2}%
}\theta _{4}\left( 0|\tau \right) \\
\theta _{1}\left( \frac{1}{2}|\tau \right) =\theta _{2}\left( 0|\tau \right)
, & \theta _{2}\left( \frac{1}{2}|\tau \right) =-\theta _{1}\left( 0|\tau
\right) \\
\theta _{3}\left( \frac{1}{2}|\tau \right) =\theta _{4}\left( 0|\tau \right)
, & \theta _{4}\left( \frac{1}{2}|\tau \right) =\theta _{3}\left( 0|\tau
\right) \\
\theta _{1}\left( \frac{\tau }{2}|\tau \right) =iq^{-\frac{1}{8}}\theta
_{4}\left( 0|\tau \right) , & \theta _{2}\left( \frac{\tau }{2}|\tau \right)
=q^{-\frac{1}{8}}\theta _{3}\left( 0|\tau \right) \\
\theta _{3}\left( \frac{\tau }{2}|\tau \right) =q^{-\frac{1}{8}}\theta
_{2}\left( 0|\tau \right) , & \theta _{4}\left( \frac{\tau }{2}|\tau \right)
=iq^{-\frac{1}{8}}\theta _{1}\left( 0|\tau \right)
\end{array}
.  \label{table1}
\end{equation}

On the basis of these properties we obtain the action of GMT on the
parameters $V$ and $\delta $ which are related to the couplings in the
boundary interactions. In this way it is possible to sort out translations
that leave the vacua unchanged and translations that change boundary
conditions by making a switching from one vacuum to another. We find that
GMT factorize into two groups acting on $V$ and $\delta $ respectively as $%
V=\lambda \frac{\pi }{2}=\frac{2\mathbf{l}+\mathbf{i}}{2}\pi $ and $\delta =%
\frac{\mathbf{i}}{2}\pi $. Only translations with $\mathbf{i}\neq 0$ (i. e.
quasi-holes) change the boundary states, i. e. the fixed points within the
boundary flow, and then act as boundary condition changing operators. For
any fixed point we find a stability group which is the subgroup of the GMT
leaving the vacuum state unchanged: it is built of any translation of
particles with $\mathbf{i}=0$ (i. e. electrons and, for $p\neq 0$, anyons).

By taking a closer look to Eq. (\ref{ZUV1}) we clearly see that the
partition function for the MR model (for $p=0$) is given by the terms in the
bracket and depends only on the parameter $V$ which is related to the
localized tunneling potential in Eqs. (\ref{SP}) or (\ref{pot1}). Both
charged and neutral components of the MR model translate together with the
same step as a result of the coupling between the two sectors due to the
parity rule. Let us notice that pure MR translations cannot be obtained
without the localized twist term, as shown in detail in the next Section. In
order to act on the MR states without modifying the last term in the
boundary partition function, Eq. (\ref{ZUV1}), it is mandatory to compensate
a $V$ translation with a translation in the $\delta $ parameter. This is a
consequence of the competition between the localized tunneling and the layer
exchange effects.

At this point we are ready to combine the action of GMT above obtained with
the modular properties of the conformal blocks (see Appendix); as a result
the parameters $V$ and $\delta $ transform as elements of the group $%
SL\left( 2,Z\right) /\Gamma _{2}$, in agreement with the conjecture about
magic points by Callan et al. \cite{maldacena}.

The results just obtained lead us to construct general GMT operators which
embody the peculiar features of our model for the quantum Hall bilayer.
Thus, in the following Section we discuss in detail such a new GMT structure
and focus on the relation between noncommutativity and non-Abelian
statistics of quasi-hole excitations.

\section{Generalized Magnetic Translations: noncommutativity and non-Abelian
statistics}

In this Section we present in detail the rich structure of Generalized
Magnetic Translations within our TM for the quantum Hall bilayer at paired
states fillings as inferred by the previous findings. This study allows us
to sort out noncommutativity and clarify its relation with non-Abelian
statistics of quasi-hole excitations. Finally a possible implementation of a
topologically protected qubit will be proposed and briefly sketched.

Our main result is that a signature of noncommutativity within our TM can be
found on GMT that become non-Abelian. In fact the quantum numbers which
label the states do not satisfy simple additive rules of composition as in
the Laughlin series but a more complex rule similar to that of spin. In
general GMT do not commute with the full chiral algebra but only with the
Virasoro one. Therefore a spectrum can be found for any of the different
vacua which correspond to different defects. In order to understand this
phenomenon let us study in detail our TM model with its boundary structure.
We will realize that noncommutativity is deeply related to the presence of
topological defects.

The whole TM can be written as $U(1)\times PF_{2}^{m}\times PF_{m}^{2}$,
where $U(1)$ is the Abelian charge/flux sector and the remaining factors
refer to neutral sector. Indeed the neutral fields can be decomposed into
two independent groups: one realizes the parafermions of the $SU(2)_{m}$
affine algebra ($PF_{m}^{2}$) and gives rise to a singlet of the twist
algebra while the second one realizes the parafermions of the $SU(m)_{2}$
algebra ($PF_{2}^{m}$) and is an irrep of the twist group. In the literature
the neutral sector was assumed to be insensible to magnetic translations due
to the neutrality \cite{MR}. Nevertheless, we will show that this is not
true in general because the breaking of the $U(1)$ pseudospin group to a
discrete subgroup implies that the residual action of the magnetic
translation group survives. Noncommutativity in the MR sub-theory $U(1)\times
PF_{2}^{m}$ of our TM arises as a result of the coupling between the charged
and the neutral component due to the $m$-ality selection rule (i. e. the
pairing phenomenon for $m=2$). In order to describe the antisymmetric part
of the full TM, instead, the coset $SU(m)_{2}/U(1)^{m-1}$must be taken into
account. The boundary interactions, Eqs. (\ref{magterm}) and (\ref{pot1}),
break the symmetry because contain operators with pseudospin $\Sigma ^{2}$
giving rise to the different fixed points of the boundary phase diagram
(with a $\Sigma _{z}$ residual symmetry) \cite{noi1}. The twist fields are
the image of noncommutativity and the fixed points correspond to different
representations with different spin. However, thanks to the modular
covariance \cite{cgm4}, we can go from a representation to the other.

\ Let us now study in detail the structure of GMT. The MR conformal blocks $%
\chi _{(\lambda ,s)}^{MR}$ (see Eqs. (\ref{mr1})-(\ref{mr3}) in the
Appendix) depend on the Abelian index $s=0,...,p$ and the spin index $%
\lambda =0,..,m$. The subgroup of GMT which stabilize the MR conformal
blocks is realized by the operators which transport symmetric electrons as
we will show in the following.

The $U(1)$ Abelian charge/flux sector is characterized by a definite $%
(q,\phi )$-charge and flux for any type of particles, which simply add
together due to the charge and flux conservation. A phase $e^{iq\phi }$ is
generated by winding one particle around another (i.e. the Abelian
Aharonov-Bohm phase factor). In the presence of the neutral component this
phase is fairly simple. We can have a different behaviour depending on the
Abelian or non-Abelian nature of the excitations. In the untwisted sector
(i.e. without $\sigma $-fields) the particles, which are anyons or
electrons, all exhibit Abelian statistics. Nevertheless the neutral
components give a contribution to the full Aharonov-Bohm phase by means of
the $Z_{m}$ charge. For the simplest $m=2$ case (quantum Hall bilayer) of
our interest in this work we encode this charge into the fermion number $F$,
counting the fermion modes, which is defined by means of $\gamma _{F}$.
Notice that in our construction we do not need to introduce cocycles for
neutral modes because the induction procedure automatically gives the
correct commutation relations for the projected fields. Nevertheless, when
we consider the charged and neutral sectors as independent it is necessary
to consider such matrices. Thus, when we decompose the $c=1$ neutral modes
into two $c=1/2$ components (see Section 2 and Appendix) two independent
Clifford algebras for fermion zero modes have to be introduced. The twisted $%
|\sigma /\mu >$ ground state is degenerate and it is possible to define a
Clifford algebra in terms of Pauli matrices, which act as $\mathbf{\Sigma }%
_{x}|\sigma /\mu >=|\mu /\sigma >;$ $\mathbf{\Sigma }_{y}|\sigma /\mu >={\pm
}i|\mu /\sigma >;$ $\ \ \mathbf{\Sigma }_{z}|\sigma /\mu >={\pm }|\sigma
/\mu >$, and of the operator $\gamma _{F}=(-1)^{F}$, which is defined in
such a way to anticommute with the fermion field, $\gamma _{F}\psi \gamma
_{F}=-\psi $, and to satisfy the property $\left( \gamma _{F}\right) ^{2}=1$%
; furthermore it has eigenvalues $\pm 1$ when acting on states with even or
odd numbers of fermion creation operators. On the above vacua the Clifford
algebra is realized in terms of the fermion modes by means of the following
operators:
\begin{eqnarray}
\gamma _{F} &=&e^{i\frac{\pi }{4}\Sigma _{z}}(-1)^{\sum \psi _{-n}\psi _{n}}%
\text{ \ \ \ and \ \ }\psi _{0}=\frac{e^{i\frac{\pi }{4}\Sigma _{x}}}{\sqrt{2%
}}(-1)^{\sum \psi _{-n}\psi _{n}}\text{\ ,\ \ twisted vacuum}
\label{clifford} \\
\gamma _{F} &=&I(-1)^{\sum \psi _{-n+1/2}\psi _{n+1/2}}\text{ ,\ \ \ \ \ \ \
\ \ \ \ \ \ \ \ \ \ \ \ \ \ \ \ \ \ \ \ \ \ \ \ \ \ \ \ \ \ \ \ \ \ \ \ \ \
\ \ \ \ \ \ untwisted vacuum}
\end{eqnarray}
in a\ $\gamma _{F}$ diagonal basis. In order to give a unified
representation of the $\gamma _{F}$ operator in the twisted as well as the
untwisted space we need to add the identity operator $I$\ in the above
definition. It acts on the layer indices space (i. e. the pseudospin space).

Within the MR model there isn't a well defined $\gamma _{F}$ in the twisted
ground state, thus we need to take vacuum states of the form $|\tilde{\sigma}%
>_{\pm }=\frac{1}{\sqrt{2}}\left( |\sigma >\pm |\mu >\right) $ while modular
invariance forces us to consider only one of these states, which corresponds
to the $\chi _{\frac{1}{16}}$ character appearing in Eq. (\ref{mr2}) (see
Appendix). In terms of fields, this would mean trading the two fields $%
\sigma $ and $\mu $ for a single field $\tilde{\sigma}_{\pm }=\frac{1}{\sqrt{%
2}}\left( \sigma \pm \mu \right) $. The fusion rule $\psi \times \sigma =\mu
$ would be replaced by $\psi \times \tilde{\sigma}_{\pm }=\pm \tilde{\sigma}%
_{\pm }$. The MR\ model contains only one of these operators ($\tilde{\sigma}%
_{+}$) which corresponds to the $\chi _{\frac{1}{16}}$ character while the
characters $\chi _{0}$ and $\chi _{\frac{1}{2}}$ have a well defined fermion
parity. As it was observed on the plane (see Section 2), the charged and the
neutral sector of MR model are not completely independent but need to
satisfy the constraint $\alpha \cdot p+l=0$ ($mod$ $2$) which is the $m$%
-ality condition (parity rule). Here such a rule is explicitly realized by
constraining the eigenvalues of the fermion parity operator upon defining
the generalized GSO projector $P=\frac{1}{2}(1-e^{i\pi \alpha \cdot p}\gamma
_{F})$. \ In this way the eigenvalues of the integer part of the neutral
translation can be related to the charged one. In order to formally extend
the definition of magnetic translations within the neutral sector to the
transport of an Abelian anyon in any of the different vacua we introduce the
couple ($a,F$) of parameters which are defined only modulo $2$. According to
the above definition $a$ is equal to $1$ for twisted ($|\tilde{\sigma}>_{+}$%
) and $0$ for untwisted vacua ($|I>$ and $|\psi >$) consistently with the
fusion rules. Transport of an anyon around another one produces a $%
(-1)^{a_{1}F_{2}-a_{2}F_{1}}$ phase and the following identification holds: $%
(a,F)=(\lambda -i-2l,l+\frac{i+\lambda }{2})$, which corresponds to the
characteristics of the Jacobi theta functions $\theta \left[
\begin{array}{c}
\frac{a}{2} \\
\frac{F}{2}
\end{array}
\right] $ within Ising characters. In conclusion, for the MR sector the
parity rule coupling between the charged and neutral sector manifests itself
as the completion of the charge/flux quantum numbers $(q,a;\phi ,F)$.

On the basis of the previous considerations the most general GMT operators
can be written as the tensor product $\mathcal{J}_{F}^{a}\otimes \mathcal{J}%
_{D}^{a}$, acting on the quantum Hall fluid and defects space respectively.
Although electrons and anyons do not exhibit non-Abelian statistics the GMT
are different for the untwisted/twisted sectors as a consequence of the
difference in the definition of $\gamma _{F}$, which is a simple identity
operator in the untwisted space but becomes a spin operator in the twisted
one. It is easy to verify that the action of \ $\mathbf{\Sigma }_{x}$ on $%
\psi _{0}$ is simply the layer exchange which takes place when the fermion\
crosses\ the defect line. When a particle encyrcles a defect, it takes a
phase $(-1)^{F}$ and changes/unchanges the pseudospin depending on the $%
\mathbf{\Sigma }/I$ operator action. The net effect of the pseudospin is to
modify the GMT bracket into an anticommutator so that the GMT algebra
becomes, for these Abelian particles, a graded algebra. We can define two
kinds of generators:

\begin{eqnarray}
\mathcal{J}_{-}^{\mathbf{a}} &=&\mathcal{J}^{\mathbf{a}}\otimes \mathbf{I},
\label{gen1} \\
\mathcal{J}_{+}^{\mathbf{a}} &=&\mathcal{J}^{\mathbf{a}}\otimes \mathbf{%
\Sigma }_{z},  \label{gen2}
\end{eqnarray}
(where $+/-$ refer to untwisted/twisted vacuum).

Here we note that the pseudospin operator which appears in $\mathcal{J}_{+}^{%
\mathbf{a}}$ is a direct consequence of the noncommutative structure of the
defect \cite{AV2}. It is made explicitly in terms of magnetic translations
operators $\mathcal{J}_{D}^{a}=U_{j_{1},j_{2}}$ given in Eq. (\ref{circles4}%
), which generate an algebra isomorphic to $SU(2)$.

A straightforward calculation tells us that $\mathcal{J}_{-}^{\mathbf{a}}$
and $\mathcal{J}_{+}^{\mathbf{a}}$ satisfy the following super-magnetic
translation algebra (within the MR sector):

\begin{eqnarray}
\left[ \mathcal{J}_{-}^{\mathbf{a}},\mathcal{J}_{-}^{\mathbf{\beta }}\right]
&=&2i\sin \left( \frac{\mathbf{s}{\times }\mathbf{s}^{\prime }}{p+1}\pi
\right) \mathcal{J}_{-}^{\mathbf{a}+\mathbf{\beta }},  \label{ssa1} \\
\left[ \mathcal{J}_{+}^{\mathbf{a}},\mathcal{J}_{-}^{\mathbf{\beta }}\right]
&=&2i\sin \left( \frac{\mathbf{s}{\times }\mathbf{s}^{\prime }}{p+1}\pi
\right) \mathcal{J}_{+}^{\mathbf{a}+\mathbf{\beta }},  \label{ssa2} \\
\left\{ \mathcal{J}_{+}^{\mathbf{a}},\mathcal{J}_{+}^{\mathbf{\beta }%
}\right\} &=&2\cos \left( \frac{\mathbf{s}{\times }\mathbf{s}^{\prime }}{p+1}%
\pi \right) \mathcal{J}_{-}^{\mathbf{a}+\mathbf{\beta }},  \label{ssa3}
\end{eqnarray}
that is the supersymmetric sine algebra (SSA).

In order to clarify the deep relationship between noncommutativity and
non-Abelian statistics in our TM let us focus on $\mathcal{J}_{D}^{a}$
operators, noncommutativity being related to the presence of topological
defects on the edge of the quantum Hall bilayer. Only on the twisted vacuum
(i. e. for $V,\delta =\frac{\pi }{2}$) the $\mathcal{J}_{D}^{a}$ realize the
spin operator $\Sigma $ while, in the usual untwisted vacuum with $V,\delta
=\pi $, there is no noncommutativity and $\mathcal{J}_{D}^{a}$ reduces to
the identity $\mathbf{I}$. The defects break the GMT symmetry so that
different backgrounds can be connected by special GMT. The breaking of the
residual symmetry of the CFT can be recognized in the condensation of the
defects. As recalled in Section 4, it is possible to identify three
different classes of defects and then three non trivial fixed points; in
particular an intermediate coupling fixed point has been found \cite{noi1},
which could be identified with the non-Fermi liquid fixed point which
characterizes the overscreened two-channel Kondo problem in a quantum
impurity context \cite{affleck1}. The $\mathcal{J}_{D}^{a}$ operators which
act on the defects space are realized by means of the $\mathrm{P}$, $\mathrm{%
Q}$ operators defined on the noncommutative torus (see Eq. (\ref{fexp13}) in
Section 3 and Ref. \cite{AV2} for details) and are taken to build up the $%
SU(m)$ boundary generators. The Clifford algebra can be realized in terms of
these operators. The non-Abelian statistics is obtained in the presence of a
$\sigma /\mu $-twist that corresponds to the defects. A tunneling phenomenon
is associated to a twist of the MR sector while a level crossing is obtained
in the presence of a twist of the pure Ising sector. The ''boundary''\ $%
SU(2) $ algebra acts on the twisted boundary conditions of the neutral
fermions. The noncommutative nature arises in the TM as a manifestation of
the vacuum degeneracy of the non-Fermi liquid fixed point, and the
interaction with the defect spin (pseudospin) is given by the Clifford
algebra. Notice that this noncommutativity is purely chiral and should be
not confused with the Aharonov-Bohm effect due to charge/flux exchange
(which is necessarily non-chiral).

Let us now define another GMT algebra which is a subalgebra of the whole GMT
algebra; it is generated by the exchange of quasi-hole excitations with
non-Abelian statistics. The GMT operators for quasi-holes are more involved
and, as shown in the previous Section, change boundary states by switching
from a twisted vacuum to an untwisted one and viceversa; as a result a
parity operator $\Omega $ with the property $\Omega ^{2}=(-1)^{F}$ must be
inserted in the definition of generators producing $\mathcal{J}%
_{D}^{a}=\Omega e^{i\frac{\pi }{4}\Sigma }$. That corresponds to the
introduction of the phase factor $e^{-i\frac{\pi }{4}}$when considering the
action on the conformal blocks \cite{AV3}. Although only closed edges are
physical, this forces us to introduce open edges as the fundamental domain
of our theory. In a string theory context the operator $\Omega $ realizes
the switching from the closed string channel to the open string one ending
on a massive $D$-brane, identified with the topological defect. Furthermore
let us notice that the defects support Majorana fermion zero modes: this
finding in our context of QHF physics parallels an analogue recent finding
by Teo and Kane in topological insulators and superconductors \cite{tk1}. It
is well known \cite{nayak} that non-Abelian statistics of quasi-hole
excitations has a $SO(2n)$ structure, typical of the so called Ising anyons.
We can realize such an algebra in our formalism by using the non-diagonal
GMT. In fact, while the product of $2n$ identical one-particle translations
are a realization of the GMT on the $2n$-particles wave functions, the
product of two independent one-particle translations can be used to realize
the braiding matrices embedded in $SO(2n)$. Let us consider here only the
neutral translations and the tensor product of $2n$ copies of such
translations. In terms of field theory this corresponds to an Ising$^{2n}$
model. Indeed it is a standard realization of $SO(2n)$ algebra which groups
the $2n$ Majorana fields into $n$ complex Dirac fields. If also one-particle
translations are added, a superextension of this algebra is obtained \cite
{isinganyons}. A representation of the braid group for $2n$ quasiholes has
dimension $2^{n-1}$and can be described as a subspace of the tensor product
of $2n$ two dimensional spaces \cite{nayak}. Each of them contains basis
vectors and the physical subspace of the tensor product is the space
generated by the vectors whose overall sign is positive. A spinor
representation of $SO(2n)\times U(1)$ lives on the tensor product space: the
$U(1)$ factor acts as a multiplicative factor, while the generators $\Sigma
_{ij}$ of $SO(2n)$ may be written in terms of the Pauli matrices $\Sigma _{i}
$. In conclusion, the braid group $\mathcal{B}_{n}$ is generated by
elementary exchanges $T_{i}$ of a quasihole $i$ and a quasihole $i+1$,
satisfying the relations:
\begin{eqnarray}
T_{i}T_{j} &=&T_{j}T_{i}\text{ \ \ \ \ \ \ \ (}|i-j|\geqslant 2\text{)}
\label{braid1} \\
T_{i}T_{i+1}T_{i} &=&T_{i+1}T_{i}T_{i+1}\text{ \ \ \ (}1\leq i\leq n-2\text{%
).}  \label{braid2}
\end{eqnarray}
They can be realized as embedded in the action of $SO(2n)\times U(1)$ as
follows: $T_{i}=\Omega e^{i\frac{\pi }{2}\Sigma _{ij}}$. The odd operators $%
T_{i}=\Omega e^{i\frac{\pi }{4}\Sigma _{i}}$ act as one-particle ones and
are identified as the GMT for a quasi-hole while the even ones, $%
T_{i}=\Omega e^{i\frac{\pi }{4}\Sigma _{i}\Sigma _{i+1}}$, are two-particle
operators. We can see that the square of a quasi-hole translation coincides
with the GMT for an electron. Thus we obtain the GMT group as the double
covering of the one-particle operators for the quasi-hole transport.

Such results can be employed in order to build up a topologically protected
qubit. Indeed a reliable implementation with a quantum Hall bilayer could be
achieved by putting a second topological defect somewhere on the edge; this
is needed in order to localize quasihole excitations. In this way we get
four quasiholes, whose positions are denoted as $\eta _{a}$, $a=1,...,4$:
quasiholes with coordinates $\eta _{1},\eta _{2}$ form the qubit while those
at $\eta _{3},\eta _{4}$ work as a tool to read and manipulate the qubit's
state \cite{qubit1}. Then it is possible to implement quantum gates by
braiding some of the quasiholes, which leads to unitary transformations in
the qubit space \cite{qubit2}.

\section{Conclusion}

In this chapter we reviewed our recent work on the physics of QHF at Jain as
well as paired states fillings in a more general context, that of a NCFT.
Indeed, when the underlying $m$-reduced CFT is put on a two-torus it appears
as the Morita dual of an Abelian NCFT. In this way noncommutativity comes
into play in our CFT description and the corresponding noncommutative torus
Lie algebra is naturally realized in terms of Generalized Magnetic
Translations (GMT). That introduces a new relationship between
noncommutative spaces and QHF and paves the way for further investigations
on the role of noncommutativity in the physics of general strongly
correlated many body systems \cite{ncmanybody}.

Paired states fillings are of main interest because of their application in
the realm of topological quantum computation. So we focused on such fillings
and, as a case study, we analyzed in detail a quantum Hall bilayer in the
presence of a localized topological defect. The system is found to be well
described by an action with two boundary interaction terms, a boundary
magnetic term and a boundary potential which, within a string theory
picture, could describe an analogue system of open strings with endpoints
finishing on $D$-branes in the presence of a background $B$-field and a
tachyonic potential. We recalled the boundary state structure corresponding
to two different boundary conditions, the periodic as well as the twisted
boundary conditions respectively, which give rise to different topological
sectors on the torus \cite{noi1}\cite{noi2}\cite{noi5}. In this context the
action of GMT operators on the boundary partition functions has been
computed and their role as boundary condition changing operators fully
evidenced. From such results we inferred the general structure of GMT in our
model and clarified the deep relation between noncommutativity and
non-Abelian statistics of quasi-hole excitations. Non-Abelian statistics of
quasi-holes is crucial for physical implementations of topological quantum
computing in QHF systems \cite{tqc1}. Work in this direction is in progress.
We also point out that noncommutativity is strictly related to the presence
of a topological defect on the edge of the bilayer system, which supports
protected Majorana fermion zero modes. That happens in close analogy with
point defects in topological insulators and superconductors, where the
existence of Majorana bound states is related to a $Z_{2}$ topological
invariant \cite{tk1}.

Our theoretical approach is peculiar in that it allows one to give a meaning
to the concept of noncommutative conformal field theory, as the Morita
equivalent version of a CFT defined on an ordinary space. Furthermore it
helps to shed new light on the relationship between noncommutativity and QHF
physics on one hand and between string and $D$-brane theory and QHF physics
on the other hand \cite{branehall2}. Recently we employed the $m$-reduction
procedure in order to describe non trivial phenomenology in different
condensed matter systems such as Josephson junction ladders and arrays \cite
{noi3}\cite{noi4}\cite{noi6}, two-dimensional fully frustrated $XY$ models
\cite{noi} and antiferromagnetic spin-$1/2$ ladders with a variety of
interactions \cite{noi7}. So, it could be interesting to investigate the
role of noncommutativity in these systems.

\section{Appendix: TM on the torus for quantum Hall bilayers at paired
states fillings}

Here we give the whole primary fields content of our TM on the torus
topology at paired state fillings by focusing on the particular case $m=2$,
which describes the physics of a quantum Hall bilayer.

On the torus, the primary fields are described in terms of the conformal
blocks (or characters) of the MR and the Ising model \cite{cgm4}. The MR
characters $\chi _{(\lambda ,s)}^{MR}$ with $\lambda =0,...2$ and $s=0,...,p$%
,\ are explicitly given by:
\begin{eqnarray}
\chi _{(0,s)}^{MR}(w|\tau ) &=&\chi _{0}(\tau )K_{2s}\left( w|\tau \right)
+\chi _{\frac{1}{2}}(\tau )K_{2(p+s)+2}\left( w|\tau \right) ,  \label{mr1}
\\
\chi _{(1,s)}^{MR}(w|\tau ) &=&\chi _{\frac{1}{16}}(\tau )\left(
K_{2s+1}\left( w|\tau \right) +K_{2(p+s)+3}\left( w|\tau \right) \right) ,
\label{mr2} \\
\chi _{(2,s)}^{MR}(w|\tau ) &=&\chi _{\frac{1}{2}}(\tau )K_{2s}\left( w|\tau
\right) +\chi _{0}(\tau )K_{2(p+s)+2}\left( w|\tau \right) .  \label{mr3}
\end{eqnarray}
They represent the field content of the $Z_{2}$ invariant $c=3/2$ \ CFT \cite
{MR} with a charged component ($K_{\alpha }(w|\tau )=\frac{1}{\eta (\tau )}%
\Theta \left[
\begin{array}{c}
\frac{\alpha }{4\left( p+1\right) } \\
0
\end{array}
\right] \left( 2\left( p+1\right) w|4\left( p+1\right) \tau \right) $) and a
neutral component ($\chi _{\beta }$, the conformal blocks of the Ising
Model).

The characters of the twisted sector are given by:
\begin{eqnarray}
\chi _{(0,s)}^{+}(w|\tau ) &=&\bar{\chi}_{\frac{1}{16}}\left( \chi
_{(0,s)}^{MR}(w|\tau )+\chi _{(2,s)}^{MR}(w|\tau )\right) ,  \label{tw1} \\
\chi _{(1,s)}^{+}(w|\tau ) &=&\left( \bar{\chi}_{0}+\bar{\chi}_{\frac{1}{2}%
}\right) \chi _{(1,s)}^{MR}(w|\tau )  \label{tw2}
\end{eqnarray}
which do not depend on the parity of $p$;
\begin{eqnarray}
\chi _{(0,s)}^{-}(w|\tau ) &=&\bar{\chi}_{\frac{1}{16}}\left( \chi
_{(0,s)}^{MR}(w|\tau )-\chi _{(2,s)}^{MR}(w|\tau )\right) ,  \label{tw3} \\
\chi _{(1,s)}^{-}(w|\tau ) &=&\left( \bar{\chi}_{0}-\bar{\chi}_{\frac{1}{2}%
}\right) \chi _{(1,s)}^{MR}(w|\tau )  \label{tw4}
\end{eqnarray}
for $p$ even, and
\begin{eqnarray}
\chi _{(0,s)}^{-}(w|\tau ) &=&\bar{\chi}_{\frac{1}{16}}\left( \chi _{0}-\chi
_{\frac{1}{2}}\right) \left( K_{2s}\left( w|\tau \right) +K_{2(p+s)+2}\left(
w|\tau \right) \right) ,  \label{tw5} \\
\chi _{(1,s)}^{-}(w|\tau ) &=&\chi _{\frac{1}{16}}\left( \bar{\chi}_{0}-\bar{%
\chi}_{\frac{1}{2}}\right) \left( K_{2s+1}\left( w|\tau \right)
-K_{2(p+s)+3}\left( w|\tau \right) \right)  \label{tw6}
\end{eqnarray}
for $p$ odd. Notice that only the symmetric combinations $\chi _{(i,s)}^{+}$
can be factorized in terms of the $c=\frac{3}{2}$ \ and $c=\frac{1}{2}$
theory. That is a consequence of the parity selection rule ($m$-ality),
which gives a gluing condition for the charged and neutral excitations.

Furthermore the characters of the untwisted sector are given by:
\begin{eqnarray}
\tilde{\chi}_{(0,s)}^{+}(w|\tau ) &=&\bar{\chi}_{0}\chi _{(0,s)}^{MR}(w|\tau
)+\bar{\chi}_{\frac{1}{2}}\chi _{(2,s)}^{MR}(w|\tau )=\chi
_{1,s}^{331}(w|\tau ),  \label{vacuum1} \\
\tilde{\chi}_{(1,s)}^{+}(w|\tau ) &=&\bar{\chi}_{0}\chi _{(2,s)}^{MR}(w|\tau
)+\bar{\chi}_{\frac{1}{2}}\chi _{(0,s)}^{MR}(w|\tau )=\chi
_{2,s}^{331}(w|\tau ), \\
\tilde{\chi}_{(0,s)}^{-}(w|\tau ) &=&\bar{\chi}_{0}\chi _{(0,s)}^{MR}(w|\tau
)-\bar{\chi}_{\frac{1}{2}}\chi _{(2,s)}^{MR}(w|\tau ),  \label{vacuum2} \\
\tilde{\chi}_{(1,s)}^{-}(w|\tau ) &=&\bar{\chi}_{0}\chi _{(2,s)}^{MR}(w|\tau
)-\bar{\chi}_{\frac{1}{2}}\chi _{(0,s)}^{MR}(w|\tau ), \\
\tilde{\chi}_{(s)}(w|\tau ) &=&\bar{\chi}_{\frac{1}{16}}\chi
_{(1,s)}^{MR}(w|\tau )=\chi _{3,s}^{331}(w|\tau )+\chi _{4,s}^{331}(w|\tau ),
\label{ut3}
\end{eqnarray}
where $\chi _{i,s}^{331}(w|\tau )$ are the characters of $331$ model \cite
{Halperin}:
\begin{eqnarray}
\chi _{1,s}^{331}(w|\tau ) &=&K^{0}\left( 0|\tau \right) K_{2s}\left( w|\tau
\right) +K^{2}\left( 0|\tau \right) K_{2(p+s)+2}\left( w|\tau \right) ,
\label{hal1} \\
\chi _{2,s}^{331}(w|\tau ) &=&K^{2}\left( 0|\tau \right) K_{2s}\left( w|\tau
\right) +K^{0}\left( 0|\tau \right) K_{2(p+s)+2}\left( w|\tau \right) ,
\label{hal2} \\
\chi _{3,s}^{331}(w|\tau ) &=&K^{1}\left( 0|\tau \right) K_{2s+1}\left(
w|\tau \right) +K^{3}\left( 0|\tau \right) K_{2(p+s)+3}\left( w|\tau \right)
,  \label{hal3} \\
\chi _{4,s}^{331}(w|\tau ) &=&K^{3}\left( 0|\tau \right) K_{2s+1}\left(
w|\tau \right) +K^{1}\left( 0|\tau \right) K_{2(p+s)+3}\left( w|\tau \right)
,  \label{hal4}
\end{eqnarray}
$K^{i}\left( 0|\tau \right) $ \ being the characters of the $c=1$ Dirac
theory. Notice that $K^{3}\left( -w|\tau \right) =K^{1}\left( w|\tau \right)
$, so that only for a balanced system the two characters can be identified
while $K^{0(2)}\left( -w|\tau \right) =K^{0(2)}\left( w|\tau \right) $. Let
us also point out that, as evidenced from Eq. (\ref{hal1}), one character of
the TM is identified with two characters of the\ $331$ model. In this way
the degeneracy of the ground state on the torus is reduced from $4\left(
p+1\right) $ to $3\left( p+1\right) $ when switching from $331$ to TM, a
clear signature of a transition from an Abelian statistics to a non-Abelian
one. Such a transition is due to the presence of two inequivalent Majorana
fermions together with the breaking of the symmetry which exchanges them. In
conclusion, while in the Halperin model the fundamental particles are Dirac
fermions with a well defined layer index, in the TM they are given in terms
of symmetric $\psi $ and antisymmetric $\bar{\psi}$ fields, that is as a
superposition of states belonging to different layers. As such, they behave
in a different way under twisted boundary conditions.

We point out that the partition function on the torus can be written as:
\begin{equation}
Z(\tau )=\frac{1}{2}\left( \sum_{s=0}^{p}2\left| \tilde{\chi}_{(s)}(0|\tau
)\right| ^{2}+Z_{untwist}^{+}(0|\tau )+Z_{untwist}^{-}(\tau
)+Z_{twist}^{+}(\tau )+Z_{twist}^{-}(\tau )\right)
\end{equation}
for $p$ even, where:
\begin{eqnarray}
Z_{untwist}^{+}(\tau )\text{{}} &=&\text{{}}\sum_{s=0}^{p}\left( \left|
\tilde{\chi}_{(0,s)}^{+}(0|\tau )\right| ^{2}+\left| \tilde{\chi}%
_{(1,s)}^{+}(0|\tau )\right| ^{2}\right) , \\
Z_{untwist}^{-}(\tau )\text{{}} &=&\text{{}}\sum_{s=0}^{p}\left( \left|
\tilde{\chi}_{(0,s)}^{-}(0|\tau )\right| ^{2}+\left| \tilde{\chi}%
_{(1,s)}^{-}(0|\tau )\right| ^{2}\right) ,
\end{eqnarray}
\begin{eqnarray}
Z_{twist}^{+}(\tau )\text{{}} &=&\text{{}}\sum_{s=0}^{p}\left( \left| \chi
_{(0,s)}^{+}(0|\tau )\right| ^{2}+\left| \chi _{(1,s)}^{+}(0|\tau )\right|
^{2}\right) , \\
Z_{twist}^{-}(\tau )\text{{}} &=&\text{{}}\sum_{s=0}^{p}\left( \left| \chi
_{(0,s)}^{-}(0|\tau )\right| ^{2}+\left| \chi _{(1,s)}^{-}(0|\tau )\right|
^{2}\right) ,
\end{eqnarray}
while for p odd we get simply:
\begin{equation}
Z(\tau )=\frac{1}{2}\left( \sum_{s=0}^{p}2\left| \tilde{\chi}_{(s)}(0|\tau
)\right| ^{2}+Z_{untwist}^{+}(0|\tau )+Z_{untwist}^{-}(\tau
)+Z_{twist}^{+}(\tau )\right) .
\end{equation}
As recalled above, the two Majorana fermions are not completely equivalent
and that reflects in the factorization of the partition function in the MR
and Ising (non-invariant) one:
\begin{equation}
Z(\tau )=Z_{MR}(\tau )Z_{\overline{I\sin g}}(\tau )
\end{equation}
where $Z_{MR}$ is the modular invariant partition function of the MR $c=3/2$
theory:
\begin{equation}
Z_{MR}(\tau )=\sum_{s=0}^{p}\left( \left| \chi _{(0,s)}^{MR}(0|\tau )\right|
^{2}+\left| \chi _{(1,s)}^{MR}(0|\tau )\right| ^{2}+\left| \chi
_{(2,s)}^{MR}(0|\tau )\right| ^{2}\right)
\end{equation}
and $Z_{\overline{I\sin g}}$ is the partition function of the Ising $c=1/2$
theory:
\begin{equation}
Z_{\overline{\text{I}\sin \text{g}}}(\tau )=\left| \bar{\chi}_{0}(\tau )%
\text{ }\right| ^{2}+\left| \bar{\chi}_{\frac{1}{2}}(\tau )\right|
^{2}+\left| \bar{\chi}_{\frac{1}{16}}(\tau )\right| ^{2}.
\end{equation}

\label{lastpage-01}

\end{document}